\documentclass{article}

\usepackage[english]{babel}
\usepackage{enumerate}
\usepackage{amsthm} 

\usepackage[a4paper,top=3cm,bottom=3cm,left=3cm,right=3cm,marginparwidth=1.75cm]{geometry}

\usepackage[numbers]{natbib}
\usepackage{pifont}

\usepackage{amsmath}

\usepackage{amsfonts}
\usepackage{bbm}
\usepackage{graphicx}
\usepackage[colorlinks=true, allcolors=blue]{hyperref}
\usepackage{mathtools}
\usepackage{breqn}
\usepackage{float}
\usepackage{adjustbox}	
\usepackage{xcolor}
\usepackage{makecell}
\numberwithin{equation}{section}

\def\E{{\mathbb{E}}}

\newcommand{\Mid}{{\ \Big|\ }}

\newcommand{\cmark}{\ding{51}}%
\newcommand{\xmark}{\ding{55}}%
\definecolor{blue0}{RGB}{0,77,153} 
\definecolor{red0}{RGB}{179,0,77} 
\definecolor{green0}{RGB}{134,219,76} 
\definecolor{gray0}{RGB}{84,97,110}

\usepackage{mathtools}
\usepackage{authblk}
\mathtoolsset{showonlyrefs}
\usepackage[normalem]{ulem}

\title{Volatility models in practice: Rough, Path-dependent or Markovian?}

\author[1]{Eduardo Abi Jaber\thanks{eduardo.abi-jaber@polytechnique.edu. The first author is grateful for the financial support from the Chaires FiME-FDD, Financial Risks, Deep Finance \& Statistics and Machine Learning and systematic methods in finance at École Polytechnique.}}
\author[2,3]{Shaun (Xiaoyuan) Li \thanks{shaunlinz02@gmail.com. \\{\quad\; {We would like to thank Christa Cuchiero, Julien Guyon, and the anonymous referee for their valuable comments.}}}}
\affil[1]{Ecole Polytechnique, CMAP}
\affil[2]{AXA Investment Managers}
\affil[3]{Université Paris 1 Panthéon-Sorbonne, CES}

\begin{document}

\maketitle

\begin{abstract}

{We present an empirical study examining several claims related to option prices in rough volatility literature using SPX options data. Our results show that rough volatility models with the parameter $H \in (0,1/2)$ are inconsistent with the global shape of SPX smiles. In particular, the at-the-money SPX skew is incompatible with the power-law shape generated by these models, which increases too fast for short maturities and decays too slowly for longer maturities. For maturities between one week and three months, rough volatility models underperform one-factor Markovian models with the same number of parameters. When extended to longer maturities, rough volatility models do not consistently outperform one-factor Markovian models. Our study identifies a non-rough path-dependent model and a two-factor Markovian model that outperform their rough counterparts in capturing SPX smiles between one week and three years, with only 3 to 4 parameters.}

\end{abstract}

\begin{description}
\item[JEL Classification:] G13, C63, G10, C45. 
\item[Keywords:] SPX options, Stochastic volatility, Pricing, Calibration, { Functional Quantization}, Neural Networks
\end{description}

\section{Introduction}\label{S:intro}
\subsection{Context}
In the realm of (rough) volatility modeling, certain claims have gained widespread acceptance within academic circles and the finance community. It has been disseminated that rough volatility models exhibit exceptional performance, seemingly reproducing the stylized facts of option prices with remarkable precision while utilizing only a limited number of parameters. Moreover, it has been argued that they outperform their traditional stochastic volatility model counterparts in capturing the essential features of volatility surfaces. These assertions have been made time and again in various research papers, 
and were often presented with a high degree of confidence.

However, the assertions about the superior performance of rough volatility models appear to rest heavily on a limited set of visual fits, often confined to specific dates or time intervals, raising concerns about their robustness over extended periods. Furthermore, there has been a notable omission in the literature concerning a fair and comprehensive comparison between rough volatility models and other models, such as the conventional Markovian stochastic volatility models and non-rough path-dependent\footnote{{ The term ``path-dependent'' in our paper refers to any model where the spot variance (or volatility) is a continuous semi-martingale but does not emit Markovian representation in finite dimension, see  Section \ref{sec:pd} for more details.}} volatility models. This absence leaves a critical gap in our understanding of the practical usefulness of rough volatility models, as the non-semimartingale and non-Markovian nature of rough volatility models have an important implementation cost that needs to be justified.

A series of recent independent empirical studies \cite{abi2022joint, delemotte2023yet,guyon2022does,romer2022empirical}, each focusing on different aspects of the volatility surface, presented evidence against the claim of ``superior performance'' attributed to rough volatility models. In particular, the following observations were made:
\begin{itemize}
    \item {The SPX at-the-money (ATM) skew does not follow a power-law as prescribed in rough volatility literature \cite{delemotte2023yet,guyon2022does};}
    \item {Rough volatility models underperform their one-factor Markovian counterparts for maturities up to three months with the same number of parameters for the joint SPX-VIX calibration problem as identified in \cite{abi2022joint,romer2022empirical}.}
\end{itemize}

Inspired by these studies, we present a new empirical study using daily SPX implied volatility surface data from CBOE spanning from 2011 to 2022. Our study integrates various aspects explored in these earlier works, and introduces new evidence and arguments to {further} challenge the perceived superiority of rough volatility models. Our empirical study is divided into two parts, with part one dedicated to options with \textit{{short}} maturities (maturities between \textit{one week and three months}), and part two on options with \textit{{short and long}} maturities (maturities between \textit{one week and three years}). All models used in our empirical study come from the same family of Bergomi-type models. These models, defined in Section \ref{S2} have (almost) the same number of parameters that can be interpreted similarly to ensure fairness of our study.

\subsection{What are rough volatility models?}
{Rough volatility models are a specific subclass of non-Markovian stochastic volatility models, where the volatility process is assumed to be a non-semimartingale process characterized by continuous paths rougher than those of standard Brownian motion. Specifically, these models employ variants of fractional Brownian motion to model the spot volatility. For instance, in \cite{bayer2016pricing}, the authors used a Riemann-Liouville fractional Brownian motion given by
\begin{equation}
X_t=\int_0^t(t-s)^{H-1 / 2}d W_s, \label{eq:louiville_frac_bm}    
\end{equation}
where $W$ is a standard Brownian motion 
to describe the spot volatility process. The parameter $H \in (0,1/2)$ here controls the local regularity of the path of $X$, i.e~$X$ is Hölder continuous of order strictly less than $H$. 
{ The Riemann-Liouville fractional Brownian motion \eqref{eq:louiville_frac_bm} with $H\in (0,1/2)$ is non-Markovian also not a semimartingale.}

{Two main empirical arguments were presented to support the use of fractional Brownian motion in volatility modeling. First, \cite{gatheral2018volatility} showed the logarithm of realized volatility time series for various equity market indices exhibits statistical ``rougher'' trajectories (lower Hölder regularity) compared to standard Brownian motion. Second, several literature such as \cite{bayer2016pricing,carr2003type,fouque2003multiscale,fukasawa2021volatility,gatheral2018volatility} argued that market ATM skew term structure exhibits power-law and explodes {for very short maturities} that is compatible with rough volatility models. However, the validity of both arguments hinges on exceedingly fine timescales that are difficult to attain in finite data sets.

In this paper, we investigate the justification for using fractional Brownian motion to model spot volatility by answering the following:
\begin{center}
    \textit{Does the assumption of non-semimartingality in rough volatility models stemming from the singularity of the fractional kernel \( t \mapsto t^{H-1/2} \) at \( t = 0 \), aligns convincingly with market-implied volatility surfaces? In other words, do the rough volatility models fit well to the volatility surface and its term structure?}
\end{center}
}

{ 
\subsection{Summary of main results}

\textbf{Fitting {\textit{short}} maturities.} Our empirical study shows that the rough Bergomi model underperforms on average compared to its one-factor Markovian counterpart with the same number of model parameters in fitting the volatility surface and the ATM skew. The one-factor Bergomi is also slightly better in predicting future volatility surface.

\textbf{Fitting  {\textit{short and long}} maturities.} The rough Bergomi model slightly, but not consistently, outperforms its one-factor Markovian counterpart with the same number of model parameters in fitting the volatility surface, the ATM skew,  and predicting future volatility surface. In addition, the rough Bergomi model clearly underperforms in all aspects compared to the two-factor (Markovian) Bergomi model with just one extra model parameter.

Our study reveals that rough volatility models are inconsistent with the global SPX volatility surface, caused by their structural limitations. In fact, the volatility surface can be much better captured by a path-dependent version of the Bergomi model, obtained by smoothing out the singularity of the fractional kernel at $0$ using a fixed value $\varepsilon>0$ and allowing the parameter $H$ to go below zero. This suggests that the problem with rough volatility models comes from their non-semimartingality nature, i.e.~the explosion of the fractional kernel as $t \xrightarrow[]{}0$ that is responsible for the roughness of the trajectory of the model's spot volatility. Moreover, we show that estimation of the roughness of the realized volatility time series is insufficient to quantify the erratic and spiky behavior that characterizes the time series of realized volatility of SPX, in line with \cite{abi2019lifting, cont2022rough,rogersthings}.}
}

\textbf{Outline of the paper}: In Section \ref{S2}, we introduce the Volterra Bergomi-type models considered in this paper and outline their key properties. Section \ref{S3} defines the performance metrics used to evaluate model performance and discusses the specifics of model calibration. In Section \ref{empirical_performance}, we present empirical results in two parts: the first focuses on {\textit{short}} maturities, while the second extends the analysis to {\textit{both short and long maturities}}. Finally, in Section \ref{S6_model_pred}, we assess the predictive capability of each model and estimate the statistical roughness of their simulated {realized} volatility paths. {Sample fits and additional calibration graphs are provided in the Appendix.}

\section{The class of Volterra Bergomi-type models}\label{S2}
The general form of the Volterra Bergomi stochastic volatility model for a (forward) stock price $S$ with spot variance $V$ is defined as
\begin{equation}\label{bergomi_family_model}
  \begin{aligned}
    dS_t &= S_t\sqrt{V_t}dB_t,\\
    V_t &= \xi_0(t) \exp{\left(X_t-\frac{1}{2}  \int_0^t K^2(s) ds \right)},\\
      X_t &= \int_0^t K(t-s) dW_s,
  \end{aligned}
  \end{equation}
with $ B=\rho W + \sqrt{1-\rho^2} W^{\perp}$, $\rho \in [-1,1]$, and $(W,W^{\perp})$ a two-dimensional Brownian motion on a risk-neutral filtered probability space $(\Omega, \mathcal F,(\mathcal F_t)_{t\geq 0}, \mathbb Q )$ satisfying the usual conditions. $X$ is a centered Gaussian process with a non-negative locally square integrable kernel $K \in L^2([0,T],\mathbb R_+)$, in particular $X_t \sim \mathcal{N}(0, \int_0^t K^2(u)du)$, for all $t \leq T$.  The deterministic input curve   $\xi_0 \in L^2([0,T],\mathbb R_{{+}})$ allows the model to match certain term structures of volatility (e.g.~the forward variance curve), since
\begin{align}\label{eq:fwdvarcalib_berg}
    \mathbb E\left[ \int_0^t V_s ds \right] = \int_0^t \xi_0(s)ds, \quad t\geq 0.
\end{align}

In this paper, we shall consider the following kernels $K$, with their corresponding model name in Table \ref{tablekernels_berg}:

\begin{table}[H]
		\centering  
            \begin{adjustbox}{width=15cm,center}
		\resizebox{\textwidth}{!}{\begin{tabular}{c c c c c} 
				\hline
				\textbf{Model name} &   ${K(t)}$ & \textbf{Domain of ${H}$} & \textbf{Semi-mart.} & \textbf{Markovian} \\
				\hline 
			\textit{rough} & $\eta t^{H-1/2}$ & $(0,1/2]$ & \textcolor{red0}{\xmark}   & \textcolor{red0}{\xmark}  \\
	\textit{path-dependent} & $\eta (t+\varepsilon)^{H-1/2}$ & $(-\infty,1/2] $ & \textcolor{blue0}{\cmark}   & \textcolor{red0}{\xmark}  \\
\textit{one-factor} & $\eta \varepsilon^{H-1/2}e^{-(1/2-H)\varepsilon^{-1} t}$ & $(-\infty,1/2]$ & \textcolor{blue0}{\cmark}   & \textcolor{blue0}{\cmark}  \\
$\textit{two-factor}$ & \makecell{$\eta \varepsilon^{H-1/2} e^{-(1/2-H)\varepsilon^{-1} t}+$\\ $\eta_{\ell} \varepsilon^{H_{\ell}-1/2} e^{-(1/2-H_{\ell})\varepsilon^{-1} t}$} & $(-\infty,1/2]$ & \textcolor{blue0}{\cmark}   & \textcolor{blue0}{\cmark} \\
    \hline 
\end{tabular}}
  \end{adjustbox}
		\caption{Different kernels $K$ and their associated Bergomi model name used throughout this paper, see Section~\ref{S:2factor} for more details.}
		\label{tablekernels_berg} 
\end{table}

To ensure comparability among all models, we fix $\varepsilon = 1/52$ {(corresponds to a timescale of 1 week)} and  $H_{\ell}=0.45$. {These values are consistent with those reported in \cite{guyon2022vix, guyon2022volatility} and ensure the existence of fast and slow factors that one usually obtains when calibrating the two-factor Bergomi model to SPX smiles}. This means that all models contain the same number of parameters to be calibrated $(\eta, \rho, H)$, except the two-factor Bergomi model which takes on an additional parameter $\eta_{\ell}$. Fixing $\varepsilon$ and $H_{\ell}$ to these values does not alter the conclusion of this paper compared to setting these parameters free.

{The particular parametrization of the one-factor and two-factor Bergomi models ensures that the parameter $H$ has a similar interpretation across all models{, see \cite{abi2023reconciling}}. This will become clearer as we introduce each Bergomi model in the section below.}

{The ATM (forward) skew $\mathcal{S}_T$ is defined as
\[
\mathcal{S}_T := \left.\frac{d\widehat \sigma (T,k)}{dk}\right\rvert_{k=0},
\]
where $\sigma (T,k)$ is the implied volatility of vanilla options calculated by inverting the Black-Scholes formula with maturity $T$ and log-moneyness $k = \log(K/F_T)$, where {$F_{T} = \E[S_T]$ is the $T-$forward price of the SPX.} For this paper, when we refer to the ATM skew, we would refer to the absolute value of $\mathcal{S}_T$.

\subsection{The rough Bergomi}
The process $X$ under the rough Bergomi model \cite{bayer2016pricing} is defined as
\begin{equation}\label{rough_bergomi}
X_t := \eta \int_0^t (t-s)^{H-1/2}dW_s,
\end{equation}
with $\eta>0$ the vol-of-vol parameter and $H\in ( 0, 1/2 ]$ that coincides with the roughness of the process $X$ and the Hurst index of its path, {i.e.~the paths of $X$ are Hölder-continuous of any order strictly less than $H$, $\mathbb{Q}$ almost surely}. For $H<1/2$, the fractional kernel $K(t)=t^{H-1/2}$ explodes as $t \xrightarrow{} 0$, so that the process $X$ is not a semi-martingale with trajectories rougher than that of standard Brownian motion. The restriction $H>0$ ensures that the kernel $K$ is locally square-integrable so that the stochastic convolution is well-defined as an Itô integral.

The rough Bergomi model produces the following ATM skew (assuming flat $\xi_0$)
{\begin{equation}\label{rough_atm_skew}
\mathcal{S}_T  \approx \frac{\rho \eta }{2(H+1/2)(H+3/2)}T^{H-1/2}
\end{equation}}
at the first-order of vol of vol, see \cite{alos2007short,bayer2016pricing,bergomi2015stochastic,fukasawa2011asymptotic}. In particular,  the skew explodes at $T\to 0$, with $H$ controlling both the skew explosive rate for small maturities and the rate of power-law decay for large maturities. {The power-law ATM skew in the form of $T^{H-1/2}$ \eqref{rough_atm_skew} is universal across all rough volatility models that use the fractional kernel to model the spot volatility/variance process \cite{alos2007short, guyon2021smile}.}

\subsection{The path-dependent Bergomi}\label{sec:pd}

Under the path-dependent Bergomi Model, the dynamics of $X$ is defined as
\begin{equation}\label{shifted_bergomi}
X_t := \eta \int_0^t (t+\varepsilon-s)^{H-1/2}dW_s.
\end{equation}
{The time shifted kernel, $K(t)=(t+\varepsilon)^{H-1/2}$, with $\varepsilon>0$ has been independently introduced over the years in \cite[Chapter 9, Equation (9.17)]{bergomi2015stochastic} and \cite{guyon2021smile}. It represents a small perturbation in the fractional kernel by $\varepsilon>0$. However, } this shift means $K(0)$ is finite, thus allowing the domain of $H \in (-\infty,1/2]$ to be extended to $-\infty$ while $K$ remains $L^2([0,T])$ integrable. The process $X$ is a continuous semi-martingale with sample paths having the same regularity as a standard Brownian motion (i.e.~Hölder-continuous of any order strictly less than $1/2$). However, $X$ is not Markovian {with respect to the filtration $(\mathcal{F}_t^W)_{t\geq 0}$ generated by the Brownian motion $W$.} To see this, we apply  Itô's formula
\begin{equation}
  \begin{aligned}
    dX_t &= \eta \Big(\int_0^t (t+\varepsilon-s)^{H-3/2}dW_s\Big)dt  + \eta \varepsilon^{H-1/2}dW_t.
    \end{aligned}\label{ito_shifted}
  \end{equation}
Next, with the help of the resolvent of the first kind of $K$, defined as the deterministic measure $L$ such that $\int_0^t K(t-s) L(ds)=1$, for all $t\geq 0$, with $L(dt)=\delta_0(dt)/K(0) + \ell (t) dt$, {where $\ell$ is a completely monotone function,}
we get 
\begin{equation}
  \begin{aligned}
    dX_t = \Big(-(1/2-H)\varepsilon^{-1} X_t + \int_0^t f(t-s) X_s ds\Big) dt + \eta \varepsilon^{H-1/2}dW_t, \end{aligned}\label{ito_shifted2}
  \end{equation}
 with
\begin{align*}
    f(t) = \frac{\varepsilon^{H-3/2}}{H-1/2}\ell (t) + \frac{1}{(H-1/2)(H-3/2)}\int_0^t (t+\varepsilon-u)^{H-5/2}\ell (u)du,
\end{align*}
see \cite[Lemma 1.2]{abi2023reconciling} for detailed  similar computations. 
$X$ is non-Markovian due to the part $\int_0^t f(t-s)X_sds$ in the drift which depends on the whole trajectory of $X$ up to time $t$.

The expression \eqref{ito_shifted2} provides essential insights into the dynamic of the process $X$: for small $t$, one expects the non-Markovian term $\int_0^t f(t-s)X_s ds$ to be negligible so that the process $X$ behaves locally like an Ornstein--Uhlenbeck process with large mean-reversion speed $(1/2-H)\varepsilon^{-1}$ and vol of vol $\varepsilon^{H-1/2}$ for small $\varepsilon$. For large $t$, the non-Markovian term $\int_0^t f(t-s)X_sds$  becomes more prominent and introduces path dependency. {Hence, the expression \eqref{ito_shifted2} justifies our use of the terminology ``path-dependent'' to describe the model, since it shows that  $X$ is a semimartingale with non-trivial path-dependency in its drift.\footnote{In the  literature,  the term ``path-dependent'' has also been used to describe models where the spot volatility (or variance)  at time $t$ depends on the past trajectory of the underlying:
$V_t = g\left(\left(S_u\right)_{u\leq t}\right)$
for some deterministic function $g$, see \cite{guyon2022volatility,hobson1998complete}. We stress that our use is different.
}}

{{Despite having a similar-looking kernel, the path-dependent Bergomi model produces very different dynamics than the rough Bergomi model}. This can be seen from its ATM skew formula \cite{guyon2021smile} below in the first-order of vol of vol, in contrast to that of the rough Bergomi model in \eqref{rough_atm_skew}
\begin{equation}
  \begin{aligned}
\mathcal{S}_T\approx\frac{\eta \rho}{2(H+1/2) T^2}\left(\frac{(T+\varepsilon)^{H+3/2}}{H+3/2}-\frac{\varepsilon^{H+3/2}}{H+3/2}-\varepsilon^{H+1/2} T\right).
    \end{aligned}\label{shift_skew_gb_expansion}
  \end{equation}

This formula shows that the global shape of the ATM skew of the path-dependent Bergomi model is more flexible than the power-law shape of the rough Bergomi model. For very short maturities, the ATM skew of the path-dependent Bergomi model approaches a finite limit {at first-order of vol of vol. Indeed, by applying the l'Hôpital's rule twice, we have
\begin{equation}\label{lim_path_dep}
    \begin{aligned}
        \lim_{T\xrightarrow{}0} & \frac{\eta \rho}{2(H+1/2) T^2}\left(\frac{(T+\varepsilon)^{H+3/2}}{H+3/2}-\frac{\varepsilon^{H+3/2}}{H+3/2}-\varepsilon^{H+1/2} T\right)\\
        =&\frac{\eta \rho}{4} \lim_{T\xrightarrow[]{}0}\frac{(T+\varepsilon)^{H+1/2} - \varepsilon^{H+1/2}}{(H+1/2)T}
        = \frac{\eta \rho}{4}  \lim_{T\xrightarrow[]{}0} {(T+\varepsilon)^{H-1/2}}\\
        =& \frac{\eta \rho \varepsilon^{H-1/2}}{4},
    \end{aligned}
\end{equation}
}
which can be made as arbitrarily large as necessary via different values of $\varepsilon$, in contrast to the blow-up to infinity in the rough Bergomi model. For longer maturities, the ATM skew of the path-dependent Bergomi model decays at a rate $\sim \rho \eta T^{H-1/2}$, with the crucial difference that $H$ can be negative, thus allowing the ATM skew to decay faster than that of rough Bergomi model.}

\subsection{The one-factor Bergomi}
{The one-factor Bergomi model introduced in \cite{bergomi2005smile} and \cite{dupire1992arbitrage} uses a standard Ornstein--Uhlenbeck process}
\begin{equation}\label{one_factor_bergomi}
X_t := \eta \varepsilon^{H-1/2} \int_0^t e^{-(1/2-H)\varepsilon^{-1}(t-s)}dW_s,
\end{equation}
where $H \in (-\infty,1/2]$. For small $\varepsilon$ and $H$, $X$ has a large mean reversion speed of order $(1/2-H)\varepsilon^{-1}$ and a large vol of vol of order $\varepsilon^{H-1/2}$ . This way of parameterizing $X$ is reminiscent of models of fast regimes in \cite{abi2023reconciling,fouque2000derivatives, mechkov2015fast} and
can be seen as a Markovian proxy of the path-dependent Bergomi model from the previous section by dropping the non-Markovian term $\int_0^t f(t-s)X_sds$ in \eqref{ito_shifted2}.

This way of parametrization allows $H$ to take on a similar interpretation to that in the path-dependent Bergomi model: the more negative the $H$, the statistically rougher the sample path of $X$ driven by larger mean reversion and vol-of-vol, see Section~\ref{S:spuriousH} below. {It also allows us to easily compare the models since they have the same parameters for calibration.}

The ATM skew produced by the one-factor Bergomi model at the first-order of vol of vol, assuming a flat $\xi_0$ is of the form \cite{bergomi2012stochastic}
\begin{align}\label{skew_one_fac}
\mathcal{S}_T \approx \frac{\rho}{2}\frac{\varepsilon^{H+1/2}\eta}{(1/2-H)T}\Big(1-\frac{(1-e^{\frac{H-1/2}{\varepsilon}T})\varepsilon}{(1/2-H)T}\Big),
\end{align} 
{which shares the same finite limit as that of the path-dependent Bergomi model {in \eqref{lim_path_dep} by applying the  l'Hôpital's rule twice:
\begin{equation}\label{lim_one_fac}
    \begin{aligned}
        \lim_{T\xrightarrow{}0} & \frac{\rho}{2}\frac{\varepsilon^{H+1/2}\eta}{(1/2-H)T}\Big(1-\frac{(1-e^{\frac{H-1/2}{\varepsilon}T})\varepsilon}{(1/2-H)T}\Big)\\
        =&\frac{\eta \rho}{4} \varepsilon^{H+1/2} \lim_{T\xrightarrow[]{}0}\frac{(1 - e^{\frac{H-1/2}{\varepsilon}T})}{(1/2-H)T}
        = \frac{\eta \rho \varepsilon^{H-1/2}}{4}  \lim_{T\xrightarrow[]{}0} {e^{\frac{H-1/2}{\varepsilon}T}}\\
        =& \frac{\eta \rho \varepsilon^{H-1/2}}{4}.
    \end{aligned}
\end{equation}
}

For small $T$, the ATM skew decay for both models is similar: {by taking the limit of the first-order derivative of \eqref{rough_atm_skew} and \eqref{lim_one_fac} at 0, the ATM skew of both models decays at the same rate as $T$ approaches zero, i.e.~}
\begin{align*}
        \lim_{T\xrightarrow{}0} \frac{d\mathcal{S}_T}{dT} = \frac{\eta\rho (H-1/2)\varepsilon^{H-3/2}}{12}.
\end{align*}

Therefore, we can expect similar model behavior between the path-dependent and one-factor Bergomi models for very short maturities, see Section \ref{short_mat_results}. On the other hand, the ATM skew decay of the one-factor Bergomi model is $\sim 1/T$ for large $T$.}

\subsection{The (under-parametrized) two-factor  Bergomi}\label{S:2factor}
The two factor Bergomi model introduced in \cite{bergomi2005smile} contains two Ornstein--Uhlenbeck factors $X^1$ and $X^2$ in the spot variance $V$:
\begin{align}
    \begin{cases}
     X_t = X_t^1 + X_t^2,\\
    X_t^1 := \eta \varepsilon^{H-1/2} \int_0^t e^{-(1/2-H)\varepsilon^{-1}(t-s)}dW_s,\\
    X_t^2 := \eta_{\ell} \varepsilon^{H_{\ell}-1/2} \int_0^t e^{-(1/2-H_{\ell})\varepsilon^{-1}(t-s)}dW_s,
    \end{cases}\label{two_factor_bergomi}
\end{align}
where the parametrization of both factors is derived in the same way as the one-factor Bergomi model above with $(H,H_{\ell}) \in (-\infty, 1/2]^2$.

In the literature, the two factors are usually driven by two correlated Brownian motions. This would require two extra parameters to model the correlation between the factors and the Brownian motion $B$ in the spot process $S$ in \eqref{bergomi_family_model}. For the sake of comparability and fairness among the models, we will use the same Brownian motion $W$ for both factors, {i.e.~$V$ is a deterministic function of a two-dimensional Markovian process $(X^1, X^2)$ with respect to the filtration $(\mathcal{F}_t^W)_{t\geq 0}$ generated by the single Brownian motion $W$,} {hence the name ``under-parametrized''}.

{By fixing $\varepsilon = 1/52, H_{\ell} = 0.45$, we induce a fast factor $X^1$ (with small $H$ and hence fast mean reversion of order $(1/2-H)\varepsilon^{-1}$), and a slow factor $X^2$ (with large value of $H_{\ell}$ and hence slow mean reversion of order $(1/2-H_{\ell})\varepsilon^{-1}$)} to mimic different scaling of volatility similar to the path-dependent Bergomi model without sacrificing the Markovian property. 

{On the ATM skew, the two factors can decouple the short and long end of the term structure, with the fast factor exerting more influence on the short end and the slow factor becoming more important as $T$ increases}. Indeed, the ATM skew assuming a flat $\xi_0$ at the first-order of vol of vol of the two-factor Bergomi model is linear in the contribution of each factor to the ATM skew \cite{bergomi2012stochastic}:
\begin{align*}
\mathcal{S}_T \approx { \frac{\rho}{2}} \Bigg(\frac{\varepsilon^{H+1/2}\eta}{(1/2-H)T}\Big(1-\frac{(1-e^{\frac{H-1/2}{\varepsilon}T})\varepsilon}{(1/2-H)T}\Big) + \frac{\varepsilon^{H_{\ell}+1/2}\eta_{\ell}}{(1/2-H_{\ell})T}\Big(1-\frac{(1-e^{\frac{H_{\ell}-1/2}{\varepsilon}T})\varepsilon}{(1/2-H_{\ell})T}\Big) \Bigg),
\end{align*}
which has a finite limit {$\frac{\rho (\eta\varepsilon^{H-1/2}+\eta_{\ell}\varepsilon^{H_{\ell}-1/2})}{4}$ when sending $T\xrightarrow{}0$} {by following similar computations as that in \eqref{lim_one_fac}}. For maturities up to three years, the two-factor Bergomi models can mimic $\sim T^{H-1/2}$ power-law decay, despite having the same asymptotic for very large $T$ as the one-factor Bergomi model.

\section{Model assessment}\label{S3}

Our empirical study involves calibrating each model described in Section \ref{S2} to the daily SPX volatility surfaces between August 2011 and September 2022 with market data purchased from the CBOE website, \url{https://datashop.cboe.com/}. In total, there are 2,807 days of SPX implied volatility surfaces. 

There is no closed-form formula for fast pricing of vanilla options under the Bergomi models {except for the one-factor Bergomi model via Fourier inversion \cite{jaber2024fourier}}. To speed up the tedious numerical optimization procedure, we rely on the generic-unified method `deep pricing with quantization hints' developed in \cite{abi2022joint}. This method allows us to price vanilla derivatives efficiently and accurately by combining Functional Quantization and Neural Networks. For detailed implementation, please refer to \cite[Section 5]{abi2022joint}.

\subsection{Treatment of the forward variance curve}\label{fvc_treatment}

We use the same forward variance curve $\xi_0(\cdot)$ across all four models inferred directly from CBOE option prices via the well-known log-contract replication formula \cite{carr2001towards}. We further assume that $\xi_0(\cdot)$ is a piece-wise constant càdlàg function as suggested by Lorenzo Bergomi himself in \cite{bergomi2015stochastic}:
\begin{align}\label{eq:pc_fvc}
\xi_0(t) = \sum_{i=1}^{N}\mathbf{1}_{t\in [T_i,T_{i+1})}\xi_i,
\end{align}
where $T_i$ are available SPX option maturities, $T_0:= 0$ and $\xi_i>0$. We can extract $\xi_i$ via
\begin{equation}
    \begin{aligned}
    (T_{i+1}-T_i) \xi_i &= 2\left(\int_0^{F_{T_{i+1}}}\frac{P(K,T_{i+1})}{K^2}dK + \int_{F_{T_{i+1}}}^{+\infty}\frac{C(K,T_{i+1})}{K^2}dK\right) \\
    &- 2\left(\int_0^{F_{T_{i}}}\frac{P(K,T_{i})}{K^2}dK + \int_{F_{T_{i}}}^{+\infty}\frac{C(K,T_{i})}{K^2}dK\right),
    \end{aligned}\label{carr_madan_formula_berg}
\end{equation}
where $C(K,T_i)$ and $P(K,T_i)$ are the market prices of vanilla call/put options with strike $K$ and maturity $T_i$. Due to the scarcity of market price for deep out-of-money options, especially for negative log moneyness, we interpolated each slice of SPX smile using a pre-determined methodology (e.g.~SSVI) and then proceeded with the computation in  \eqref{carr_madan_formula_berg}. Note we are only using the interpolated surface to estimate $\xi_0(t)$. The actual calibration is performed using the CBOE data.

\subsection{Calibration performance metrics}\label{perf_criteria}

For model evaluation, {we examine the accuracy of the model fit to the global SPX smiles, the ATM skew, and the model's prediction of future SPX smiles}.

\subsubsection{Global fit of the implied volatility surface}\label{calib_prod}

To test the global fit of the SPX smiles, each model is calibrated to minimize the error function between the model implied volatility surface and that of SPX over the set of model
parameters, denoted collectively as $\Theta$. For the rough, path-dependent, and one-factor Bergomi models, $\Theta = \{\eta, \rho, H\}$. For the two-factor Bergomi model, $\Theta = \{\eta, \rho, H, \eta_{\ell}\}$. We chose the Root Mean Square Error (RMSE) as the error function $\mathcal{J}(\Theta)$
\begin{equation}
  \begin{aligned}
    & \mathcal{J}(\Theta) := \sqrt{ \frac{1}{\vert \mathcal{I} \vert} \sum_{(i,j)\in \mathcal{I}} \bigg( \widehat \sigma_{i,j}^{mid} - \widehat \sigma_{i,j}(\Theta) \bigg)^2},
  \end{aligned}\label{calibration_error_berg}
\end{equation}
with $\widehat\sigma_{i,j}^{mid}$ the SPX mid implied volatility with maturity $T_i$ and strike $K_{j}$, and $\widehat\sigma_{i,j}(\Theta)$ the model implied volatility. The set of available SPX implied volatility data for different maturities and strikes is captured by the index set $\mathcal{I}$, with $\vert \mathcal{I} \vert$ denoting the total number of available data points.

Due to the availability of market data as well as the stability of the implied volatility estimator, we use the following log-moneyness range shown in Table \ref{log_moneyness_table}. 
{\begin{table}[H]
{
		\centering  
            \begin{adjustbox}{width=6cm,center}
		\resizebox{\textwidth}{!}{\begin{tabular}{c c} 
	\hline
	\textbf{Maturities $T$} &   \textbf{log moneyness range}\\
				\hline 
	$<$ 2 weeks & $[-0.15,0.03]$ \\
	$<$ 1 month & $[-0.25,0.03]$ \\
        $<$ 2 months & $[-0.3,0.04]$ \\
	$<$ 3 months & $[-0.4,0.15]$ \\
	$<$ 6 months & $[-0.6,0.15]$ \\
	$<$ 1 year & $[-0.8,0.2]$ \\
	$\geq$ 1 year & $[-1.5,0.3]$ \\ 
 \hline 
		\end{tabular}}
  \end{adjustbox}
		\caption{Log moneyness range for different maturities for model calibration to the global SPX volatility surface.}
		\label{log_moneyness_table} 
}
\end{table}}

\subsubsection{Fit of the implied volatility ATM skew}\label{S:fitskew}
To evaluate the model fit of the ATM skew, each model is calibrated to the SPX ATM skew by minimizing the error between model implied volatility and the mid-SPX implied volatility over a smaller range of log moneyness $k \in [-0.05, 0.03]$ across all maturities, and the error between the log of model ATM skew and log of SPX ATM skew
\begin{align}\label{eq:rmse_atm_skew}
    &\min_{\Theta} \Bigg\{\sqrt{ \frac{1}{\vert \mathcal{I} \vert}\sum_{(i,j)\in\mathcal{I}} \Big(\widehat \sigma_{i,j}^{mid} - \widehat \sigma_{i,j}(\Theta) \Big)^2 + \frac{1}{\vert \mathcal{M} \vert} \sum_{m\in\mathcal{M}}  \Big(\log\mathcal{\widehat S}_{T_{m}}^{mkt}- \log \mathcal{\widehat S}_{T_{m}}(\Theta) \Big)^2} \Bigg\},
\end{align}
where $\mathcal{\widehat S}_{T_{m}}^{mkt}$ is the SPX ATM skew at maturity $T_m$, with $\mathcal{\widehat S}_{T_{m}}(\Theta)$ is the model ATM skew. The set of available market ATM skew for different maturities is captured by the index set $\mathcal{M}$.

Thanks to the mesh-free nature of our deep pricing method that combines Functional Quantization and NN, we can calibrate each model directly to the SPX ATM skew without resorting to the approximation formula of the Bergomi-Guyon expansion as is the case in \cite{delemotte2023yet, guyon2022does}.

To compute each model's ATM skew, we first compute the model's implied volatility near the money and then use the central finite difference at $k=0$. The SPX ATM skew is computed by fitting a {polynomial} of order 3 locally near the money and then taking the first-order derivative at $k=0$. We checked our results to ensure no over- or under-fitting.

\subsubsection{Prediction quality (parameter stability)}\label{pred_desc}

{To evaluate the prediction quality of a model, we perform the following experiment: for each trading day, we take the calibrated parameters $\Theta^*$ obtained under Section \ref{calib_prod}, and keep it fixed for the next 20 working days. Next, we take the daily market forward variance curve $\xi_0(\cdot)$ extracted as per \eqref{fvc_treatment} and compute the error function \eqref{calibration_error_berg} for each of the next 20 working days with {the same} $\Theta^*$. Ideally, a robust parametric model should not require frequent re-calibration. Therefore, this performance metric can also be seen as a test of the stability of model parameters.

This performance metric is similar to the one described in \cite[Section 4.2]{romer2022empirical}. However, the treatment of $\xi_0(\cdot)$ in \cite{romer2022empirical} is not entirely consistent across every model: the $\xi_0({\cdot})$ for the Markovian Heston model is modeled by only three parameters $(V_0, \kappa, V_{\infty})$, where the 
rough Bergomi and rough Heston models received preferential treatment by employing piece-wise constant function between maturities that offers greater flexibility in controlling the overall level of the model implied volatility.
}

\section{Empirical results: fitting SPX smiles}
\label{empirical_performance}

\subsection{\textit{Short} maturities (one week to three months)}\label{short_mat_results}

We now compare the calibration performance between rough, path-dependent, and one-factor Bergomi models. Recall these models share the same parameters $(\eta, \rho, H)$ to be calibrated.

Figure \ref{fig:rmse_3mths} shows the time series of daily calibration RMSE of each model fitted to the global SPX smiles as per \eqref{calibration_error_berg}. For ease of comparison, we show the monthly rolling average RMSE. The RMSE for the path-dependent Bergomi model is almost always below that of the rough Bergomi model, while the one-factor Bergomi model also outperforms the rough Bergomi model {63 percent of the time}. The summary of statistics in {Table \ref{tb:stats_3mth} shows that the rough Bergomi model scores the highest error across the board.} Some sample fits are provided in Appendix \ref{sample_fit_3mths_appendix}.

  \begin{figure}[H]
    \centering    \includegraphics[width=0.8\textwidth]{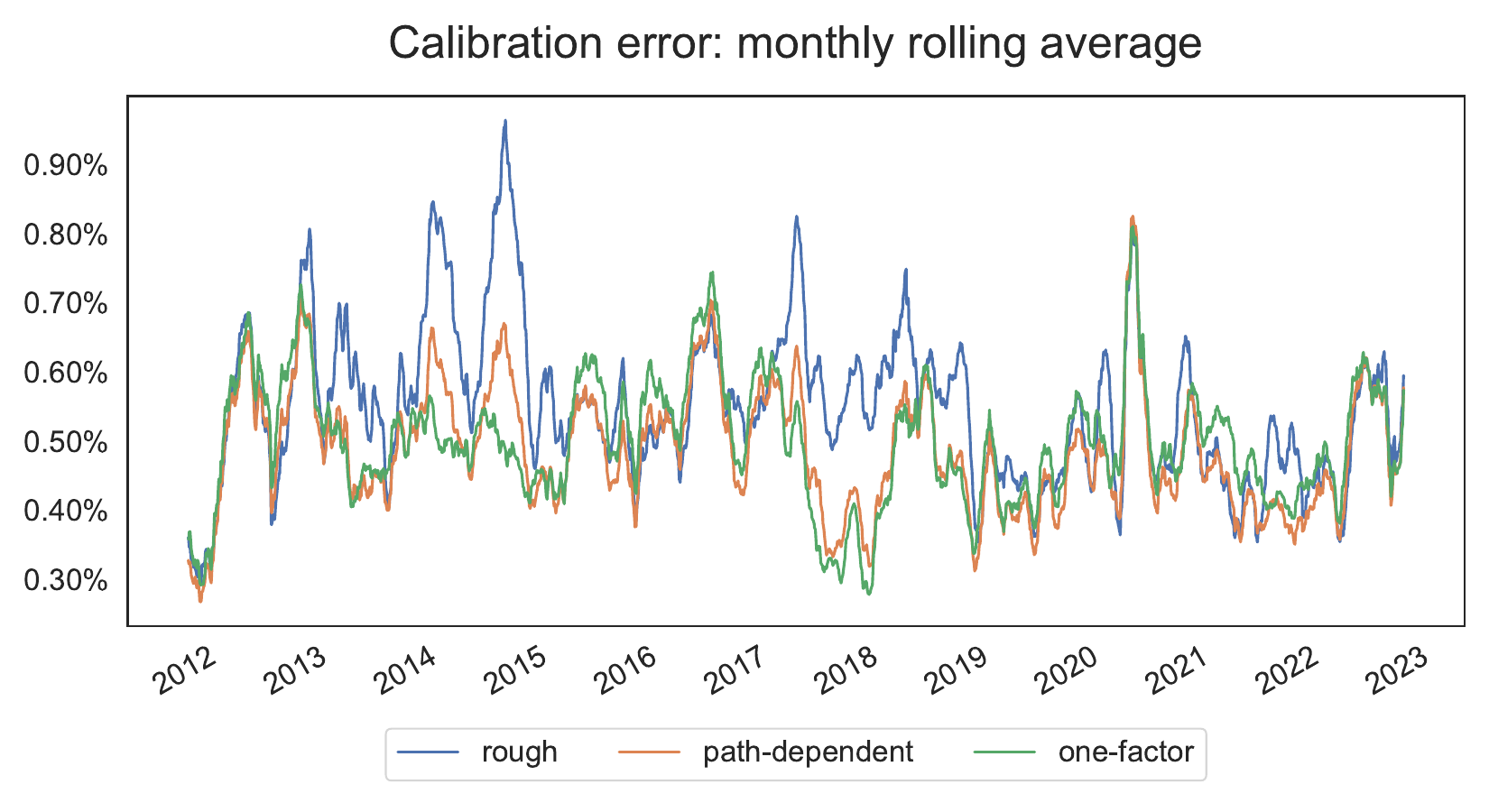}%
    \caption{Time series of monthly rolling average of calibration RMSE between different Bergomi models for \textit{short} maturities.}
    \label{fig:rmse_3mths}
  \end{figure}

{\begin{table}[H]
   \centering  
            \begin{adjustbox}{width=12cm,center}
		\resizebox{\textwidth}{!}{\begin{tabular}{c c c c c c c c} 
	\hline
	\textbf{Model}&\textbf{mean}&\textbf{std}&\textbf{min}&\textbf{5\%}&\textbf{50\%}&\textbf{95\%}&\textbf{max}\\
				\hline 
	\textbf{rough} & 0.0055&	0.0018&	0.0011&	0.0030&	0.0053&	0.0087&	0.0198 \\
 	\textbf{path-dependent} & \textbf{0.0049}&	0.0016&	\textbf{0.0009}&	\textbf{0.0027}&	\textbf{0.0047}&	\textbf{0.0075}&	0.0189\\
 	\textbf{one-factor} & 0.0050&	\textbf{0.0015}&	0.0011&	0.0028&	0.0048&	0.0076&	\textbf{0.0162}\\
 \hline 
		\end{tabular}}
  \end{adjustbox}
\vspace{0.02cm}
        \caption{Summary of statistics of the calibration RMSE for \textit{short} maturities. The lowest value for each statistical measure is in \textbf{bold}.}
        \label{tb:stats_3mth} 
\end{table}}

Figure \ref{fig:fit_skews_3mths} shows the model fit to the average SPX ATM skew and the log average SPX ATM skew. Even though no model seems to be able to perfectly capture the entire term structure of the average ATM skew {for \textit{short} maturities}, the fit of the rough Bergomi model is evidently worse than the one-factor and path-dependent Bergomi models, characterized by an overly steep ATM skew around one week of maturity and a much too rapid decay straight after. 

The above results suggest that rough volatility models are inconsistent with SPX volatility surface for {\textit{short} maturities}. In addition, SPX ATM skew cannot be adequately explained by a single power-law, which agrees with the results in \cite{delemotte2023yet, guyon2022does}. We refer the reader to Appendix \ref{rmse_fit_skew_appendix_3m} for more graphs and statistics on the calibration of the ATM skew for \textit{short} maturities.}
  \begin{figure}[H]
    \centering
\includegraphics[width=0.5\textwidth]{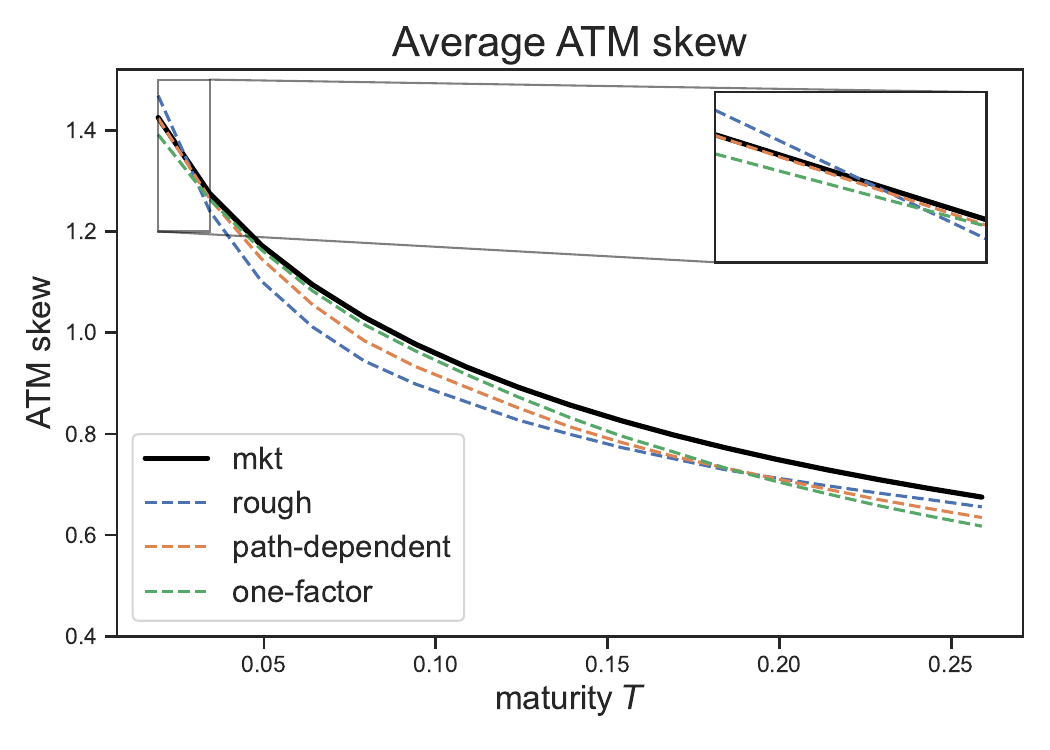}%
\includegraphics[width=0.5\textwidth]{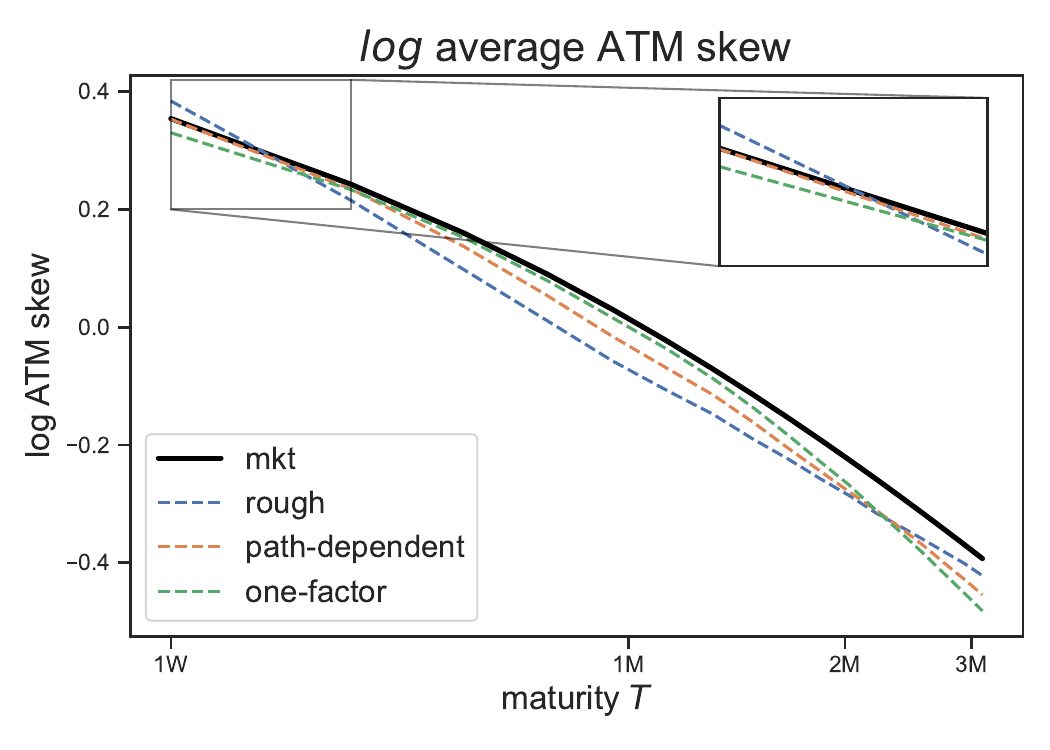}
\caption{Term structure of average SPX ATM skew (left) and log average SPX ATM skew in log-scale (right) fitted by different Bergomi models {for \textit{short} maturities}, with zoom-in for maturities between one week and two weeks.} \label{fig:fit_skews_3mths}
  \end{figure}
\vspace{-0.6cm}

Contrary to what rough volatility literature claims, {we found no evidence supporting that rough volatility models fit better than their Markovian counterparts for \textit{short} maturities of the SPX smile}. Instead, our study shows that the one-factor Bergomi outperforms the rough Bergomi model.

{For the calibrated parameters}, Figure \ref{fig:calib_rho_3mth} {in Appendix \ref{rmse_fit_vs_appendix_3m}} shows the calibrated $\rho$ for the rough Bergomi model tends to saturate near $-1$ (and even more so for {\textit{short and long}} maturities, see Section \ref{long_mat_results}): this is a known structural issue of the rough Bergomi model, see \cite{forde2020rough,horvath2021deep, romer2022empirical}. {Figure \ref{fig:calib_H_3mth} in Appendix \ref{rmse_fit_vs_appendix_3m} shows that the calibrated $H$ for the path-dependent and one-factor Bergomi models are almost always negative. 

\subsection{\textit{Short and long} maturities (one week to three years)}\label{long_mat_results}

We now compare the empirical results for {\textit{short and long}} maturities for the rough, path-dependent, one-factor, and under-parametrized two-factor Bergomi models.

Figure \ref{fig:rmse_2year} and Table \ref{tb:stats_3year} show the two-factor Bergomi model outperforms all other models for the global fit of the implied volatility surface, followed by the path-dependent Bergomi model. The performance of the rough Bergomi model is noticeably unsatisfactory. Even if one could argue that its performance is slightly better than that of the one-factor Bergomi model (57.4 percent of the time), the rough Bergomi model consistently underperformed the one-factor Bergomi model between 2017 and 2019. More sample fits are provided in  Appendix \ref{sample_fit_3years_appendix}.
  \begin{figure}[H]
    \centering    \includegraphics[width=0.8\textwidth]{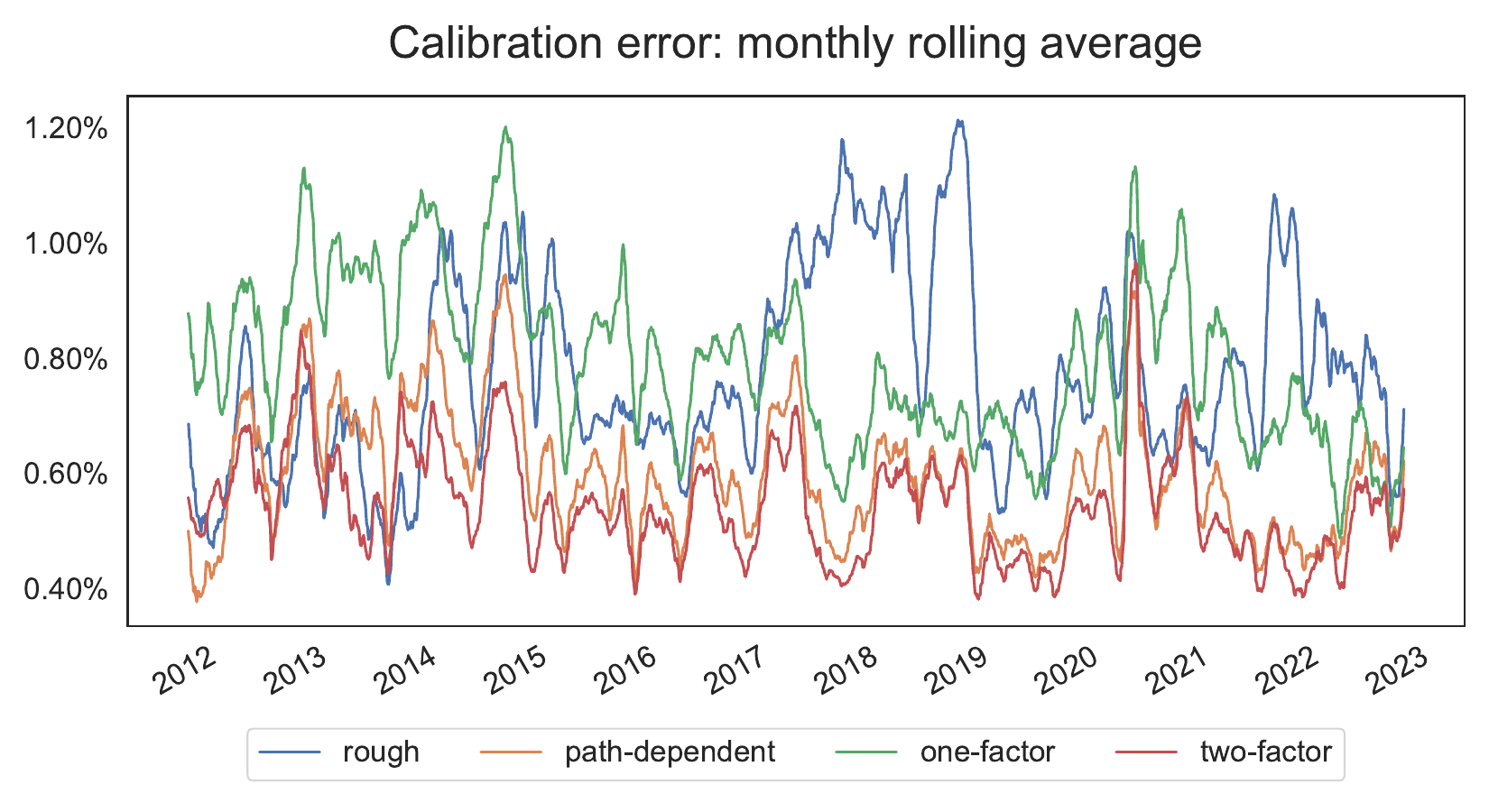}%
    \caption{Time series of monthly rolling average of calibration RMSE between different Bergomi models for \textit{short and long} maturities.}
    \label{fig:rmse_2year}
  \end{figure}

{\begin{table}[H]
   \centering  
            \begin{adjustbox}{width=12cm,center}
		\resizebox{\textwidth}{!}{\begin{tabular}{c c c c c c c c} 
	\hline
	\textbf{Model}&\textbf{mean}&\textbf{std}&\textbf{min}&\textbf{5\%}&\textbf{50\%}&\textbf{95\%}&\textbf{max}\\
				\hline 
	\textbf{rough} & 0.0077&	0.0021&	0.0029&	0.0048&	0.0074&	0.0115&	\textbf{0.0209} \\
 	\textbf{path-dependent} & 0.0059&	0.0016&	\textbf{0.0023}&	0.0037&	0.0058&	0.0087&	0.0215\\
 	\textbf{one-factor} & 0.0079&	0.0019&	0.0033&	0.0051&	0.0078&	0.0112&	0.0225\\
 	\textbf{two-factor} & \textbf{0.0054}&	\textbf{0.0014}&	0.0024&	\textbf{0.0035}&	\textbf{0.0053}&	\textbf{0.0078}&	0.0220\\
 \hline 
		\end{tabular}}
  \end{adjustbox}
\vspace{0.02cm}
        \caption{Summary of Statistics of the calibration RMSE for \textit{short and long} maturities. The lowest error for each statistical measure is in \textbf{bold}.}
        \label{tb:stats_3year} 
\end{table}}

Figure \ref{fig:fit_skews_3years} shows that the two-factor and path-dependent Bergomi models generally produce decent, though imperfect, fits to the full-term structure of the SPX ATM skew. The rough Bergomi and the one-factor Bergomi models produce equally bad fits and are inconsistent with the general shape of SPX ATM skew. In line with the results in \cite{delemotte2023yet,guyon2022does}, our study challenges the assumption used by rough volatility advocates that the SPX ATM skew follows a {single} power-law, which increases too fast in the short end and does not decay as fast in the long end when compared to SPX data. {We refer the reader to Appendix \ref{rmse_fit_skew_appendix_3y} for additional graphs and statistics on the calibration of the ATM skew.} {We also provide sample fits to SPX ATM skew on two dates in Appendix \ref{sample_fit_skew_3years_appendix}. On both days, the path-dependent and two-factor Bergomi models are flexible enough to capture the overall shape of the ATM skew term structure compared to the rough and one-factor Bergomi models.}

  \begin{figure}[H]
    \centering
    \includegraphics[width=0.5\textwidth]{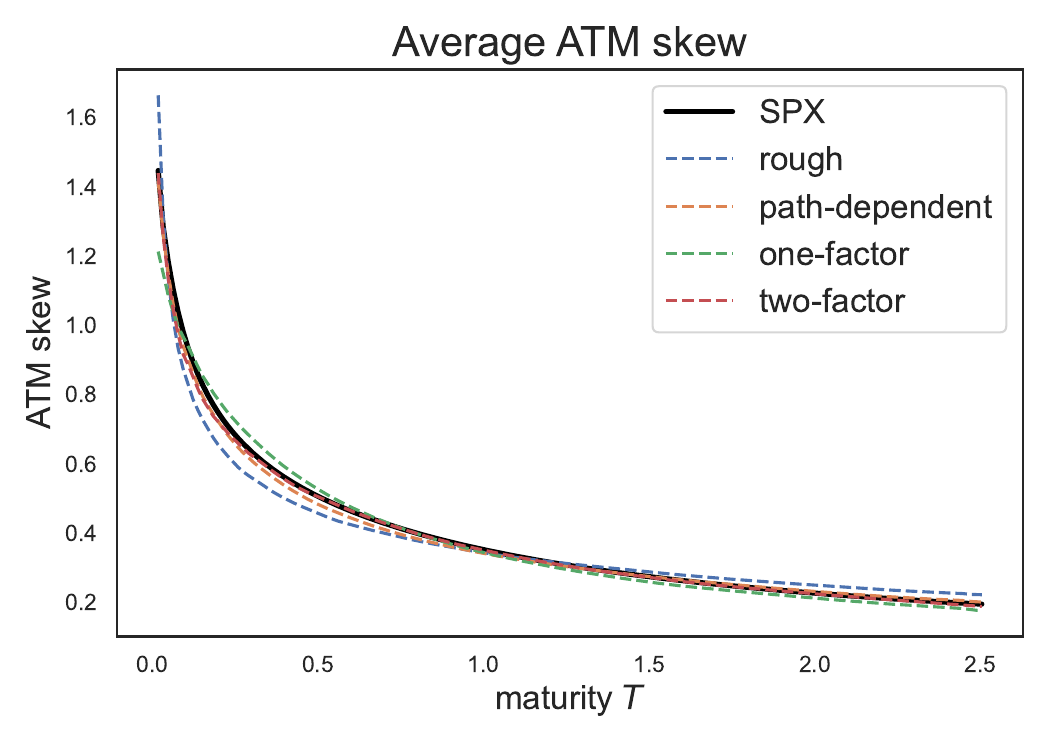}%
    \includegraphics[width=0.5\textwidth]{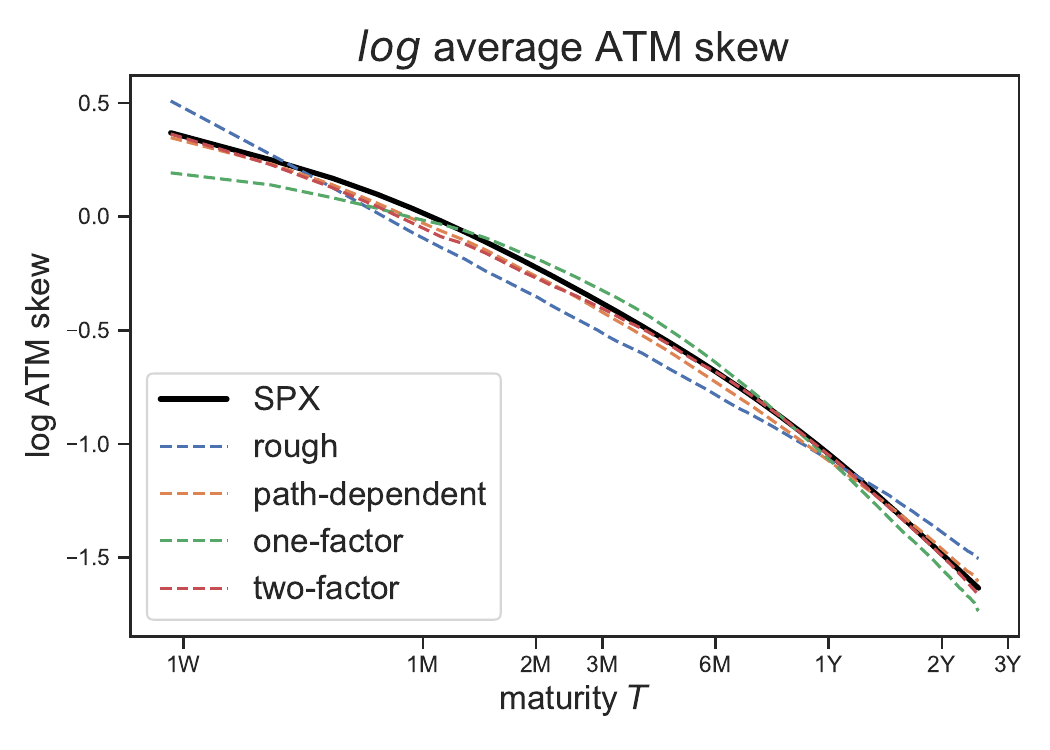}
    \caption{Term structure of average SPX ATM skew (left) and log average SPX ATM skew in log-scale (right) fitted by different Bergomi models {for \textit{short and} maturities}.}  
    \label{fig:fit_skews_3years}
  \end{figure}
  \vspace{-0.6cm}
  
Overall, our empirical study suggests that rough volatility models are inconsistent with SPX volatility surface for both {\textit{short and long} maturities.} and go against the narrative that rough volatility models outperform their Markvovian parts in fitting the full SPX smiles} {: the rough and one-factor Bergomi models exhibit comparable performance, and it is sufficient to add a second Markovian factor with an extra parameter $\eta_{\ell}$ to outperform the rough Bergomi model. At this stage, one might consider incorporating a second rough process into the rough Bergomi model. However, as noted in \cite{delemotte2023yet,romer2022empirical}, this does not significantly improve its fit. } 

Figure \ref{fig:calib_rho} {in Appendix \ref{rmse_fit_vs_appendix_3y}} shows that the calibrated $\rho$ for the rough Bergomi model tends to saturate at $-1$, whereas the calibrated $\rho$ for all other models are less saturated and tend to move together. Figure \ref{fig:calib_H} in Appendix \ref{rmse_fit_vs_appendix_3y} shows that the calibrated $H$ for the path-dependent and two-factor Bergomi models are almost always negative. In addition, the calibrated $H$ for the path-dependent Bergomi model is about 0.4 higher than the calibrated $H$ for {\textit{short}} maturities. The increase in $H$ is necessary to help the model take care of the larger maturities, since the more negative the $H$, the faster the ATM skew decay for larger $T$.

\subsection{Interpretation of the { the negative \texorpdfstring{$H$}{Lg} parameter}}\label{negative_H_sec} 

\begin{figure}[H]
    \centering    \includegraphics[width=0.5\textwidth]{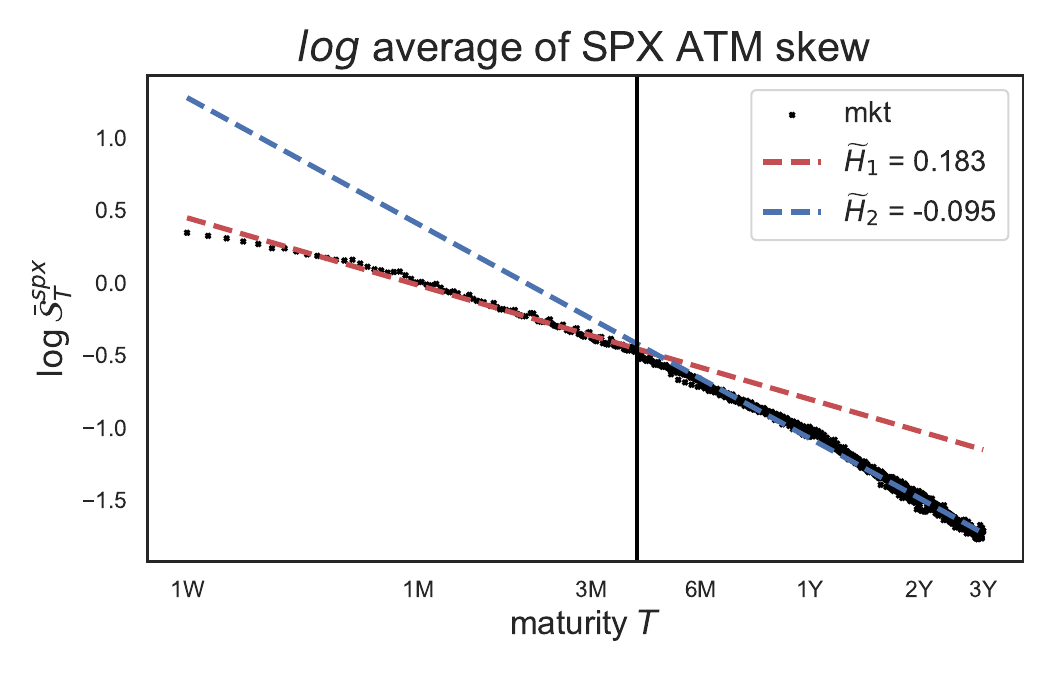}%
    \caption{Fitting the $\log \mathcal{\bar{S}}_T^{spx}$ with two linear regressions on $\log T$, with the first linear regression performed on $T\in [\text{1W, 4M})$ and the second linear regression performed on $T\in [\text{4M, 3Y}]$. The slope of each linear regression represents estimated \text{$\widetilde H$} at different timescales.}
    \label{fig:two_power_law_fit}
  \end{figure}

In two recent independent studies \cite{delemotte2023yet,guyon2022does}, two different power-laws were used to fit the full SPX ATM skew. We perform the same experiment by fitting two power-laws of the form $cT^{{\widetilde H}-1/2}$ on the average SPX ATM skew using linear regression. Specifically, we fit one power-law on $(\log T,\log \mathcal{\bar{S}}_T^{spx})$, with $\mathcal{\bar{S}}_T^{spx}$ the average SPX ATM skew over the period for $T<\tau$ and another power-law for $T\geq \tau$ and infer the two different values of ${\widetilde H}$. In Figure \ref{fig:two_power_law_fit}, $\tau$ is chosen to be 4 months based on the highest average $R^2$ value of the two linear regressions among all possible values of $\tau$ between one week and three years.

\begin{figure}[H]
    \centering    \includegraphics[width=0.5\textwidth]{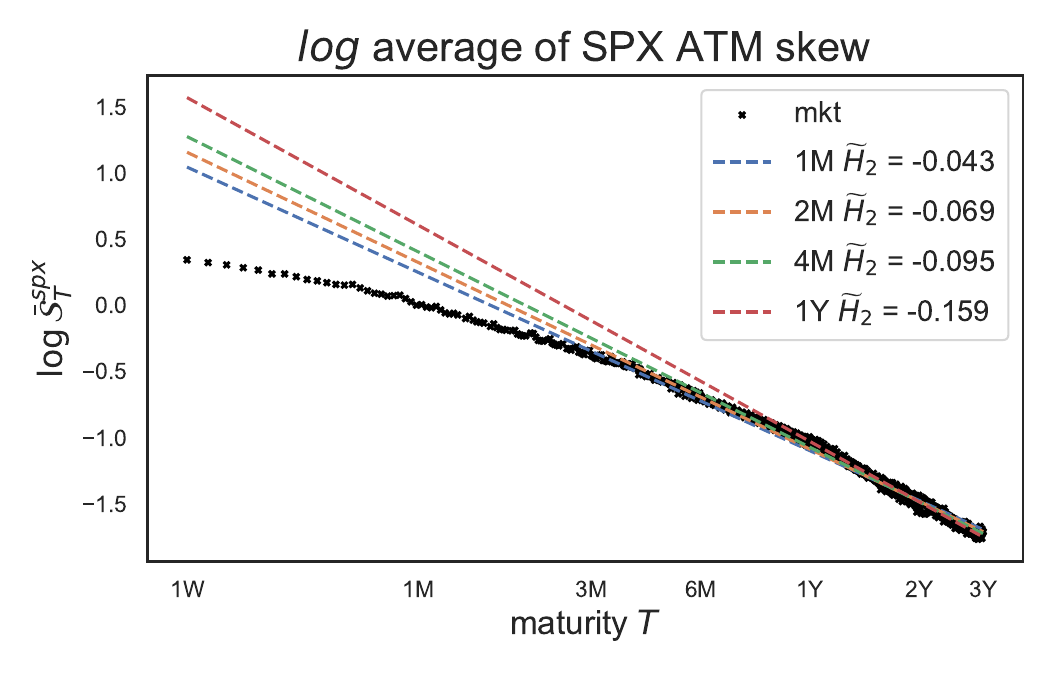}%
    \caption{Fitting the $\log \mathcal{\bar{S}}_T^{spx}$ with linear regression on $\log T$ for $T \in [\tau, \text{3Y}]$ with $\tau \in \{\text{1M,2M,4M,1Y}\}$. The slope of each linear regression represents estimated ${\widetilde H_2}$ at different timescales.}
    \label{fig:diff_H}
  \end{figure}

The decent fit of the blue dash line in Figure \ref{fig:two_power_law_fit} suggests that long-term ATM skew can be well-captured by a power-law but with a negative ${\widetilde H_{2}}=-0.095$. 
We now show what happens to the estimation of ${\widetilde H_{2}}$ by moving the cut-off time $\tau$ between one month and one year in Figure \ref{fig:diff_H}. The estimated ${\widetilde H_{2}}$ becomes increasingly negative as $\tau$ grows. This indicates that, on average, the long-term SPX ATM skew decays as a power-law $T^{\widetilde H_2 -1/2}$ with $\widetilde H_2<0$, which is steeper than previously reported $T^{-1/2}$ in \cite{bergomi2015stochastic,fouque2004maturity,fukasawa2021volatility,gatheral2008consistent}.

\begin{figure}[H]
    \centering    \includegraphics[width=0.8\textwidth]{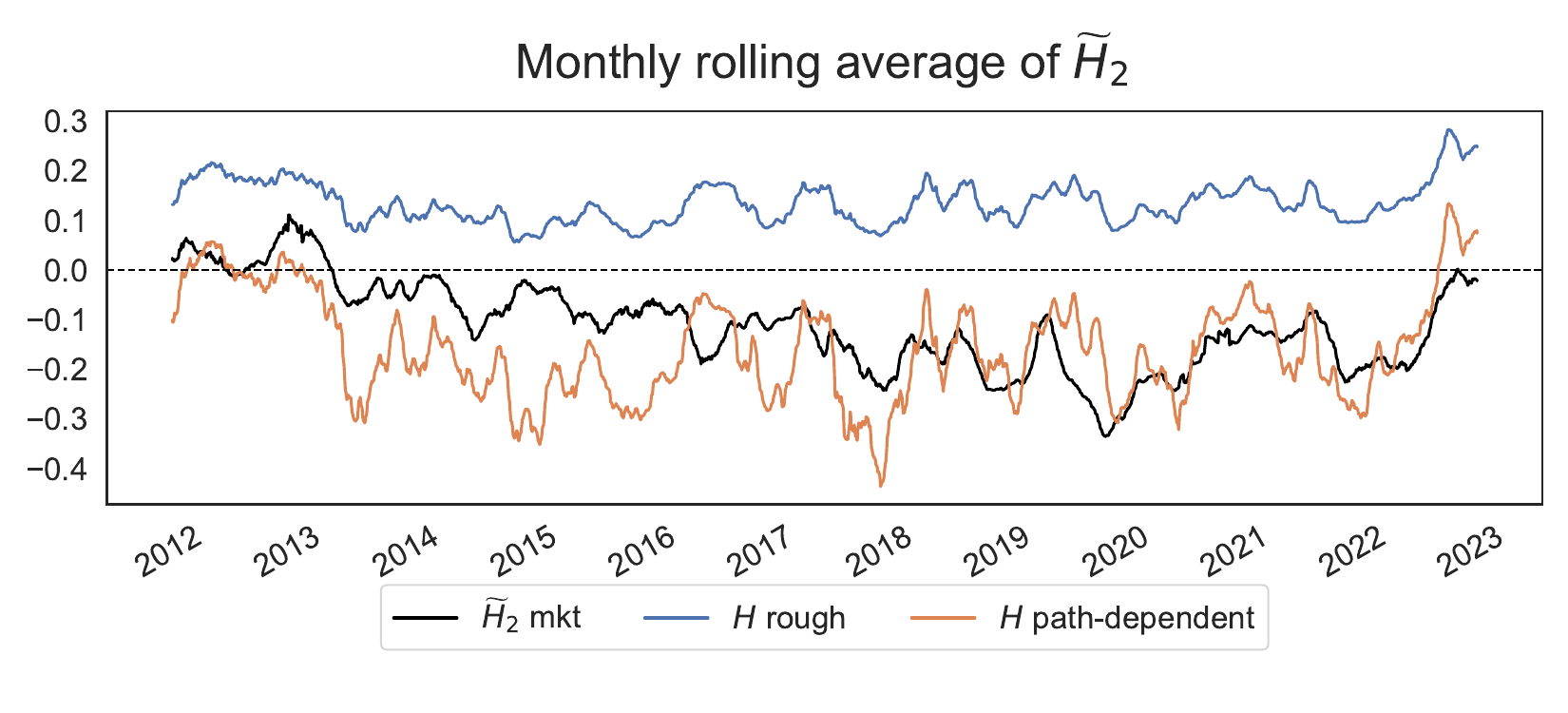}%
    \caption{Time series of regressed ${\widetilde H_2}$ estimated by fitting the daily log SPX ATM skew against $\log T$ for $T\in [\text{1Y, 3Y}]$, compared to the calibrated value of $H$ of the rough and path-dependent Bergomi models.}
    \label{fig:H_evol_mkt}
  \end{figure}

{To further validate the negative values of ${\widetilde H_2}$ observed in Figures \ref{fig:two_power_law_fit} and \ref{fig:diff_H}, we perform a linear regression by fitting the daily SPX log ATM skew term structure against $\log T$, for $T \in [\text{1Y, 3Y}]$, and plot the time series of regressed ${\widetilde H_2}$ in Figure \ref{fig:H_evol_mkt}, showing that the estimated market ${\widetilde H_2}$ remains almost always negative, except for the period from 2012 to mid-2013. For comparison, we also plot the calibrated values of $H$ from the rough and path-dependent Bergomi models, which were fitted to the ATM skew for \textit{short and long} maturities. Recall that the rough Bergomi model can produce a power-law, however, it can only decay as fast as $T^{-1/2}$ due to the $H>0$. This limitation does not apply to the path-dependent Bergomi model, which can produce a power-law decay of the form $T^{H-1/2}$
with the parameter $H$ allowed to be negative. This partially explains the superior performance of the path-dependent Bergomi model in Figure \ref{fig:fit_skews_3years}.}

\section{Empirical results: predicting future SPX smiles}\label{S6_model_pred}

\begin{figure}[H]
    \centering    \includegraphics[width=0.8\textwidth]{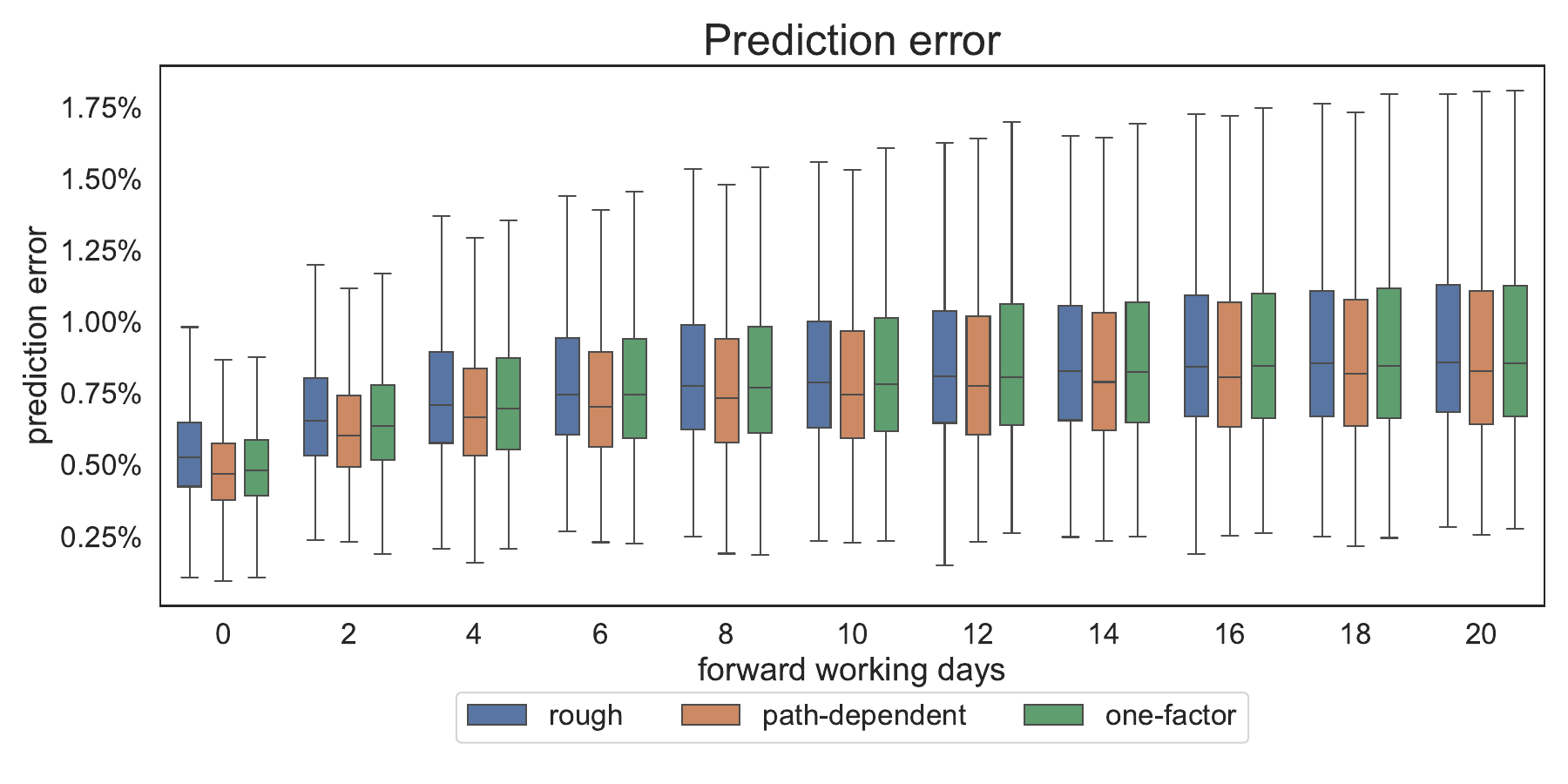}%
\caption{Box plot of prediction RMSE between different Bergomi models for \textit{short} maturities. Different values on the box plot represent 25, 50, and 75 quantiles, while the whiskers are calculated as 1.5 multiplied by the inter-quartile range away from 25 and 75 quantiles.}
\label{fig:rmse_pred_3mths}
  \end{figure}

We now look at how well each model can predict the future SPX volatility surface, { using the methodology} described in Section \ref{pred_desc}.
The box plot in Figure \ref{fig:rmse_pred_3mths} shows the distribution of RMSE of prediction quality for the period August 2011 to September 2022 when the model parameters are calibrated to \textit{short} maturities. The path-dependent Bergomi model scores the best performance. The one-factor Bergomi model performs better than the rough Bergomi model for the first six forward business days and shares similar performance afterward.

\begin{figure}[H]
    \centering    \includegraphics[width=0.8\textwidth]{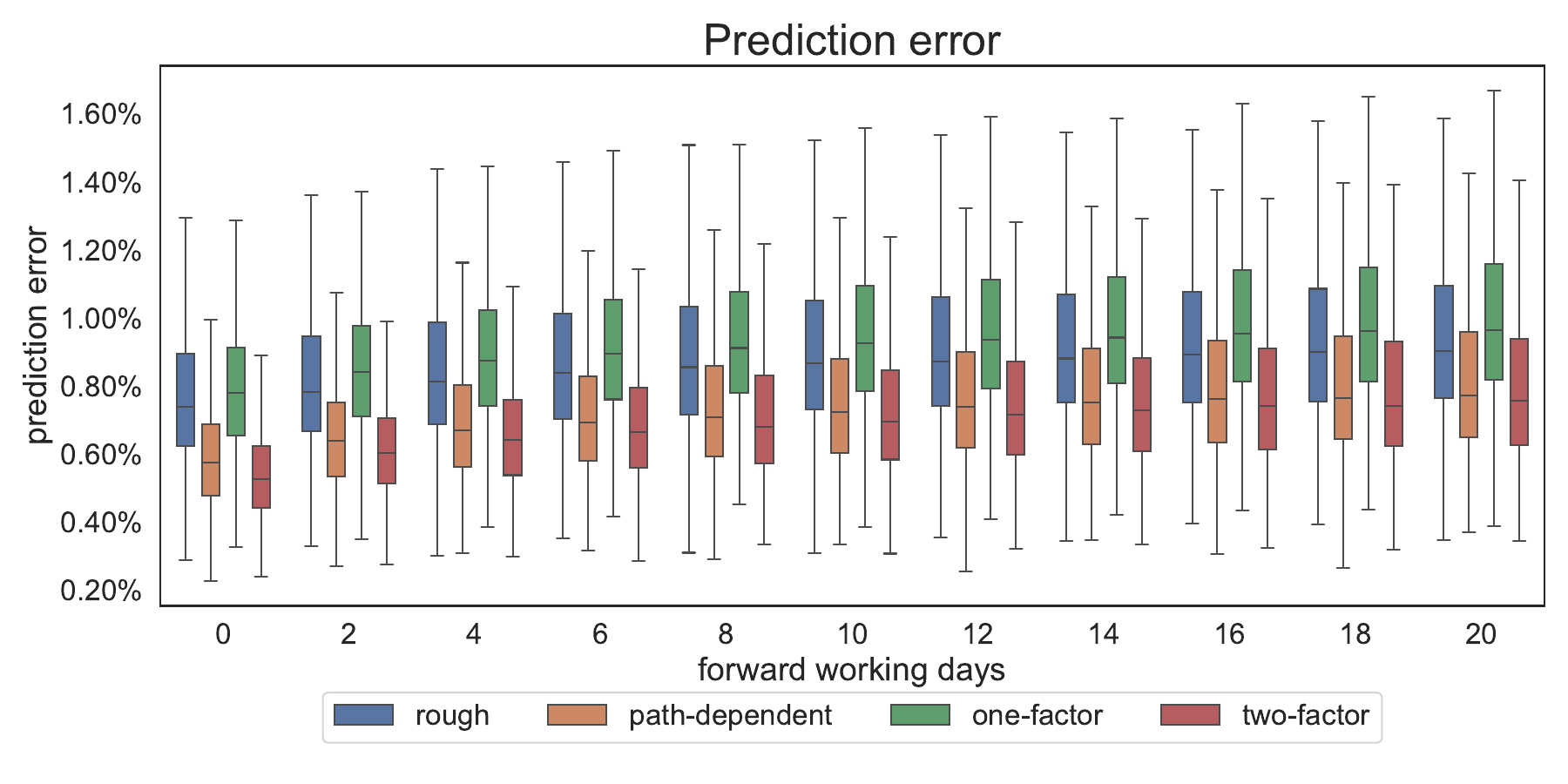}%
\caption{Box plot of prediction RMSE between different Bergomi models for \textit{short and long} maturities. Different values on the box plot represent 25, 50, and 75 quantiles, while the whiskers are calculated as 1.5 multiplied by the inter-quartile range away from 25 and 75 quantiles.}
    \label{fig:rmse_pred_2years}
  \end{figure}

The box plot in Figure \ref{fig:rmse_pred_2years} shows the distribution of RMSE of prediction quality for the period August 2011 to September 2022 when the model parameters are calibrated to \textit{short and long} maturities. The two-factor Bergomi model is the best among all the models in predicting future volatility surface, followed by the path-dependent Bergomi model. The rough Bergomi model slightly outperforms the one-factor model but is inadequate compared to its path-dependent and (two-factor) Markovian counterparts.

\section{Spurious roughness of realized volatility}\label{S:spuriousH}

{ Is roughness in the spot volatility process a necessary condition for the roughness observed in the realized volatility time series? \cite{rogersthings} showed that an OU process with high mean reversion and vol of vol, coupled with another OU process with lower mean version and vol of vol provides a good fit to the SPX realized volatility time series. In \cite[Section 4.2]{abi2019lifting}, it is shown that a finite superposition of semimartingales with fast mean reversion and volatility trick the human eye as well as statistical estimators of the Hurst index leading to spurious roughness on a wide range of timescales.   The authors in \cite{cont2022rough} demonstrated that, regardless of the roughness of the spot volatility, the realized volatility time series always exhibits rough trajectories, corresponding to an estimated Hurst index that is significantly smaller than 0.5. They went even further and suggested that the observed roughness in the realized volatility time series arises primarily from the estimation error of spot volatility using realized volatility.

To complement these results, we attempt to estimate the statistical roughness of the realized volatility of each model using the same methodology as in \cite{gatheral2018volatility}, and compare these estimates with the theoretical roughness of each model's spot volatility processes.} First, we simulate the trajectories of the spot volatility $\sqrt{V}$ and $\log S$ for each model, with a time step size of 5 minutes for $T=10$ years. Next, we compute the estimated daily $RV$ defined as
\[
\textit{RV}_{t_i} := \sqrt{\sum_{i=1}^{n} \Big(\log(S_{t_i}/S_{t_{i-1}})\Big)^2},
\]
with $t_n-t_0 = \text{1 day}$ and then compute the empirical q-variation, defined as
\[
m(q, \Delta) := \frac{1}{N}\sum_{k=1}^N \Mid \textit{RV}_{k\Delta} - \textit{RV}_{(k-1)\Delta} \Mid^q
\]
for different values of $q$ and timescale $\Delta$. To remain as close as to the experiments performed in \cite{gatheral2018volatility}, we choose $q \in \{0.5, 1, 1.5, 2, 3 \}$, with timescale $\Delta = 1, 2, \ldots, 50$ days. Specifically, we perform linear regression on ($\log \Delta, \log m(q, \Delta))$ to estimate its slope $\zeta_q$ and then plot $(q, \zeta_q)$, with the slope of the graph $(q, \zeta_q)$ being the estimated Hurst index {$\widehat H$} of the simulated trajectory of $\textit{RV}$.

To simulate spot volatility $\sqrt{V}$, we first calibrate each model to the implied volatility surface as of 23 October 2017 for \textit{short} maturities, and then perform Monte Carlo simulation using the calibrated parameters in Table \ref{calib_param_example} and the same seed.

{\begin{table}[H]
{
		\centering  
            \begin{adjustbox}{width=8cm,center}
		\resizebox{\textwidth}{!}{\begin{tabular}{c c} 
	\hline
	\textbf{Model} &   \textbf{Calibrated parameters}\\
    \hline 
	rough & $\eta = 1.28, \rho =-0.940, H =0.079$ \\
	path-dependent & $\eta = 0.0256, \rho =-0.688, H =-1.276$ \\
	one-factor & $\eta = 0.756, \rho =-0.684, H =-0.364$ \\
        two-factor & $\eta = 0.430, \eta_{\ell} = 0.984, \rho =-0.685, H =-0.497$ \\
 \hline 
		\end{tabular}}
  \end{adjustbox}
		\caption{Calibrated model parameters for \textit{short} maturities of the implied volatility surface as of 23 October 2017.}
	\label{calib_param_example} 
}
\end{table}}

  \begin{figure}[H]
    \centering
    \includegraphics[width=0.26\textwidth]{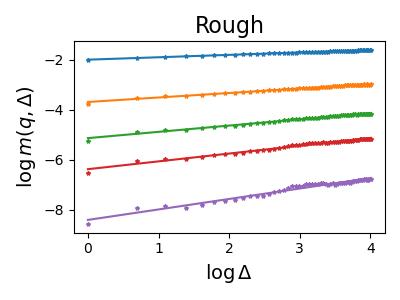}%
    \includegraphics[width=0.26\textwidth]{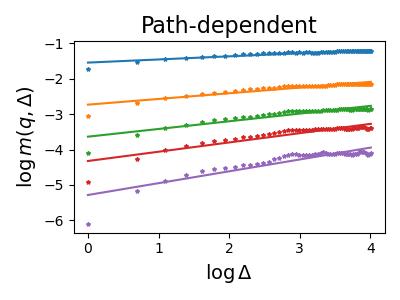}%
    \includegraphics[width=0.26\textwidth]{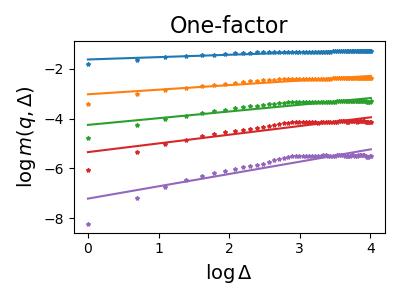}%
    \includegraphics[width=0.26\textwidth]{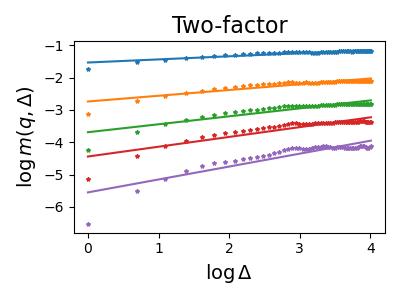}
    \caption{Log-log plot of $q$-variations of a sample path of the realized volatility for different models.}  
    \label{fig:log_log_plot}
  \end{figure}

\begin{figure}[H]
  \centering
  \includegraphics[width=0.26\textwidth]{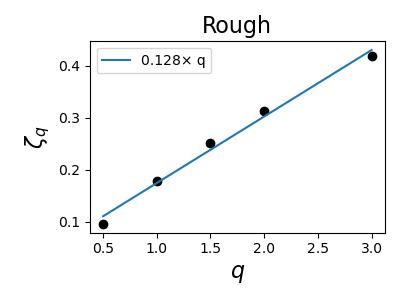}%
  \includegraphics[width=0.26\textwidth]{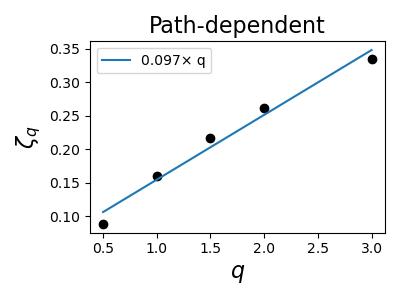}%
  \includegraphics[width=0.26\textwidth]{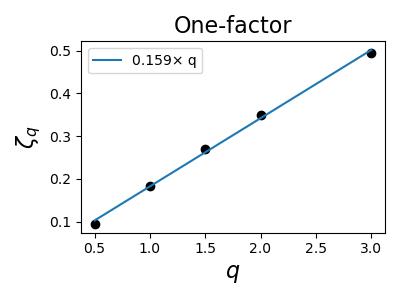}%
  \includegraphics[width=0.26\textwidth]{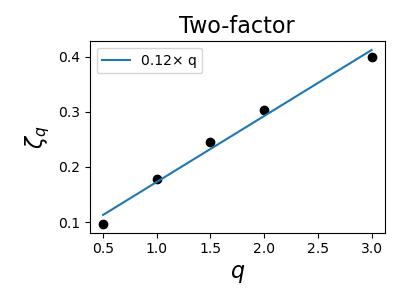}
  \caption{Plot of $\zeta_q$ against $q$, with the slope being the estimated Hurst index $\widehat H$ for different models.}  
  \label{fig:q_zeta_plot}
\end{figure}

  \begin{figure}[H]
    \centering
    \includegraphics[width=0.26\textwidth]{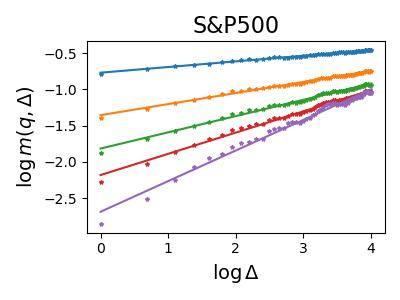}%
    \includegraphics[width=0.26\textwidth]{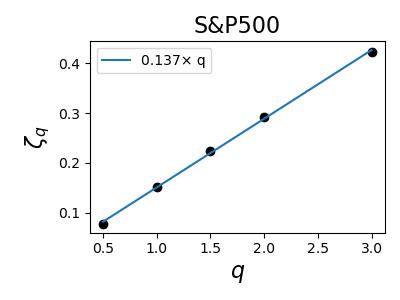}
    \caption{LHS: Log-log plot of $q$-variations of the S\&P500 realized volatility time series between 2007 and 2017. RHS: Plot of $\zeta_q$ against $q$, with the slope being the estimated Hurst index $\widehat H$ for the S\&P500 realized volatility time series.}  
    \label{fig:sp500_market_plots}
  \end{figure}

Figure \ref{fig:log_log_plot} and \ref{fig:q_zeta_plot} show that the estimate for the Hurst index of the rough Bergomi model is $\widehat H \approx 0.13$ compared to the calibrated value of $H = 0.08$ for the spot volatility. For the other models, the estimated Hurst index $\widehat H$ of $RV$ returns a value between 0.10 and 0.16, compared to the theoretical Hurst index of 0.5 of their spot volatility.  {Figure \ref{fig:sp500_market_plots} shows the estimated $\widehat H \approx 0.14$ of the actual S\&P500 realized volatility over ten years between 2007 and 2017 based on the data from Oxford-Man Institute of Quantitative Finance Realized Library.} Our study confirms the results in \cite{abi2019lifting,rogersthings}: roughness of the {realized} volatility process does not imply that the underlying process has to be rough, and is in general not an adequate criterion to be used for model selection.

{\section{Conclusion}}

{In this paper, we presented a comprehensive empirical study using the Volterra Bergomi model class and daily option data from 2011 to 2022. The result of our empirical study provides clear evidence that, despite their additional complexity, rough volatility models underperform their Markovian counterparts when calibrated to SPX smiles. Our findings reinforce previous empirical studies highlighting the inconsistency of rough volatility models with the market ATM skew, and extend these results to the entire SPX volatility surface. We also demonstrated that statistical estimation based on a single realized volatility trajectory is inadequate to determine the true dynamics of market volatility.}

\appendix

\section{Sample fits of SPX smiles}

\subsection{\textit{Short} maturities}\label{sample_fit_3mths_appendix}

\underline{July 03, 2013}

  \begin{figure}[H]
    \centering
    \includegraphics[scale=0.52]{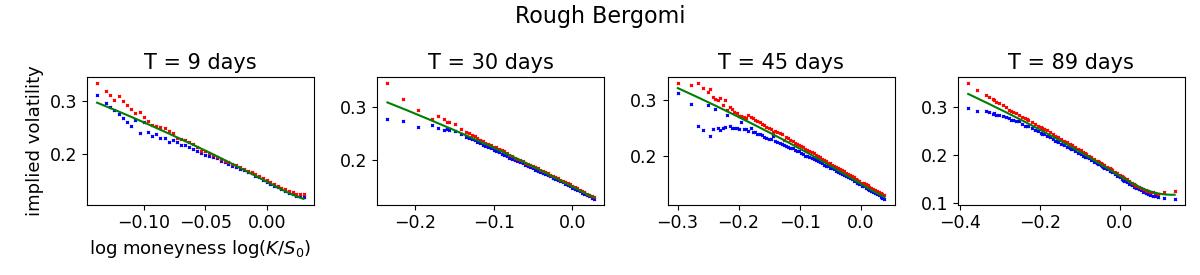}
    \includegraphics[scale=0.52]{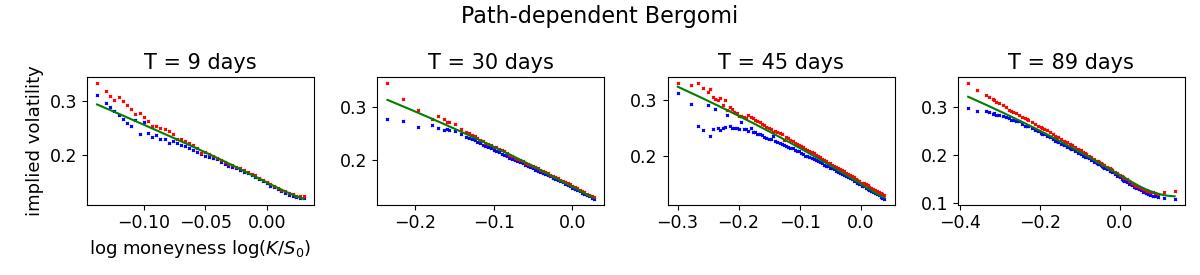}
    \includegraphics[scale=0.52]{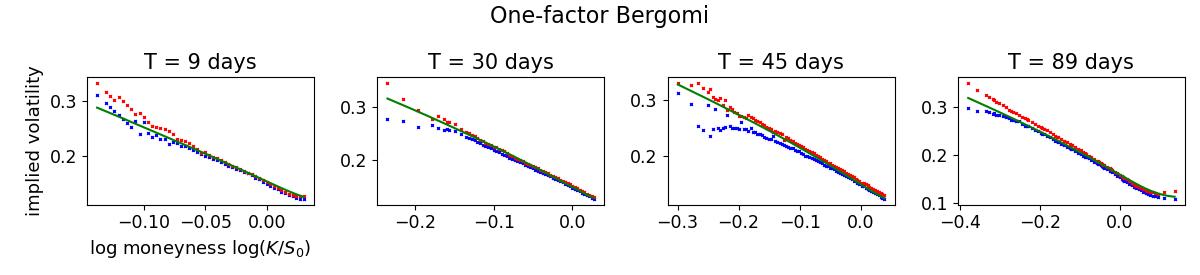}
    \caption{SPX smiles (bid/ask in blue/red) on 3 July 2013 calibrated by different Bergomi models (green lines).}
    \label{fig:calib_3mth_2013_08_14}
  \end{figure}

\underline{October 23, 2017}

  \begin{figure}[H]
    \centering

    \includegraphics[scale=0.52]{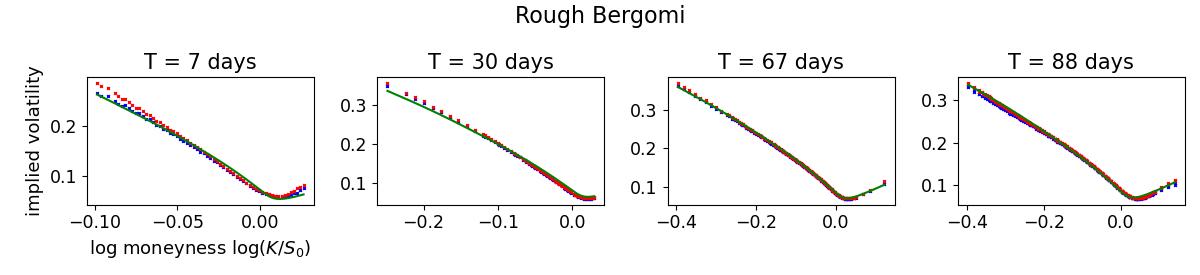}
    \includegraphics[scale=0.52]{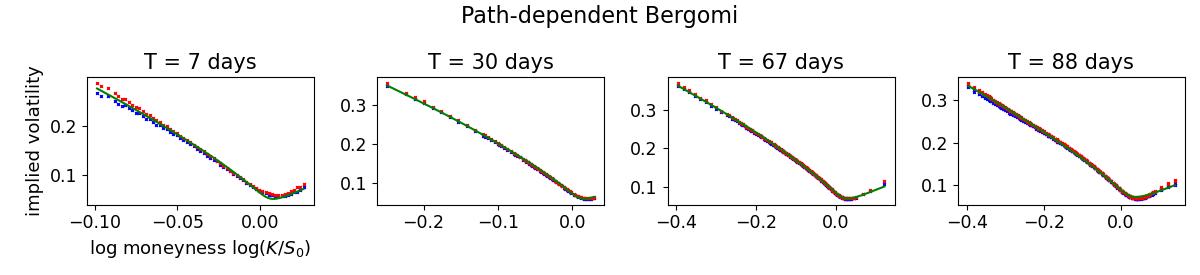}
    \includegraphics[scale=0.52]{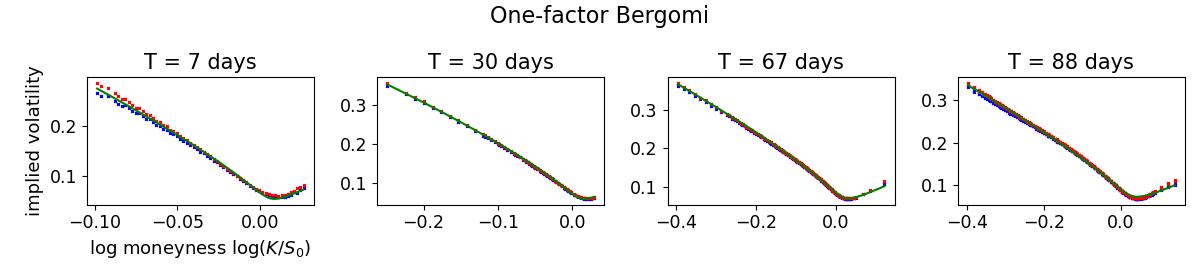}
    \caption{SPX smiles (bid/ask in blue/red) on 23 October 2017 calibrated by different Bergomi models  (green lines).}
    \label{fig:calib_3mth_2017_11_22}
  \end{figure}

\subsection{\textit{Short and long} maturities}
\label{sample_fit_3years_appendix}

\underline{July 03, 2013}

  \begin{figure}[H]
    \centering    \includegraphics[width=1.01\textwidth]{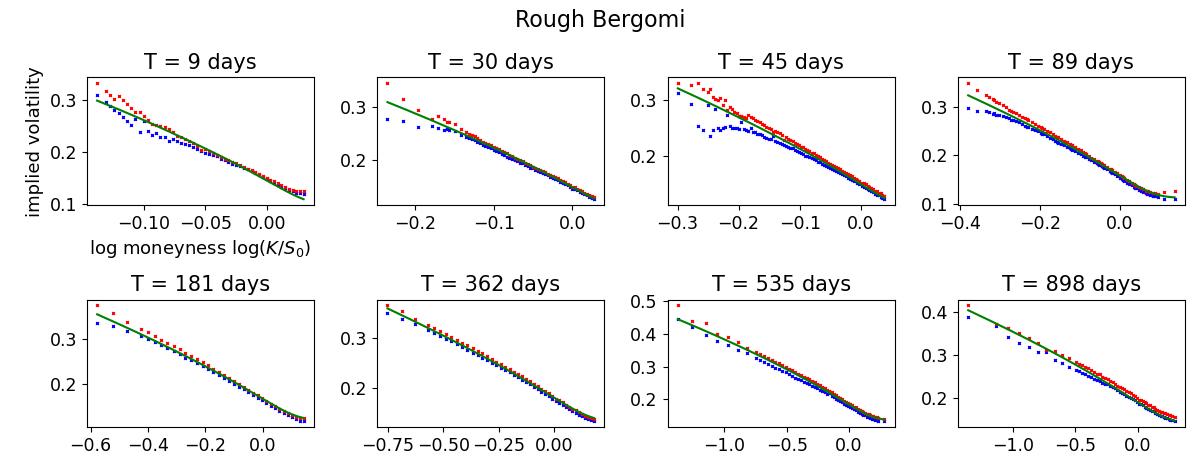}%
    \caption{SPX smiles (bid/ask in blue/red) on 3 July 2013 calibrated by \textbf{rough} Bergomi models (green lines).}
  \end{figure}

  \begin{figure}[H]
    \centering    \includegraphics[width=1.01\textwidth]{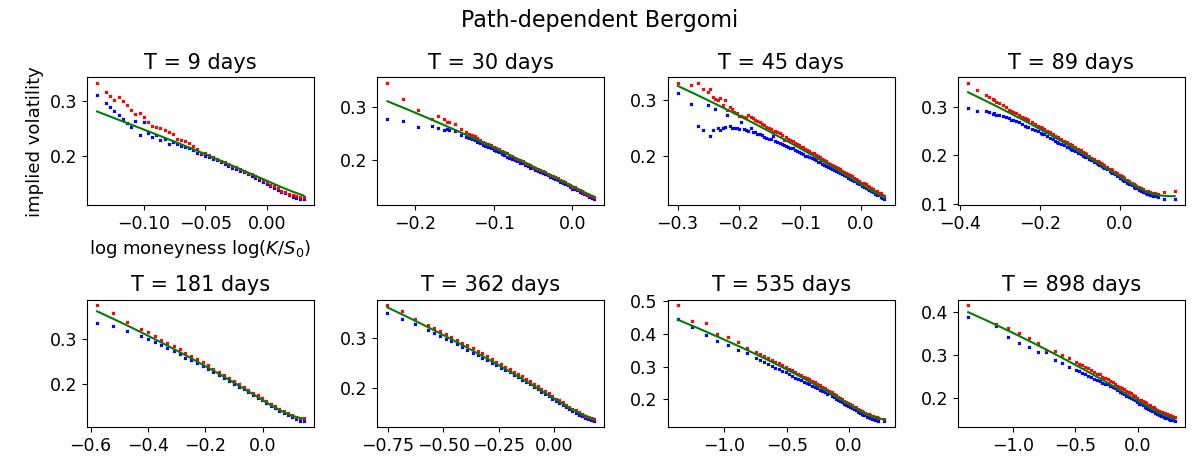}%
    \caption{SPX smiles (bid/ask in blue/red) on 3 July 2013 calibrated by \textbf{path-dependent} Bergomi models (green lines).}
  \end{figure}

  \begin{figure}[H]
    \centering    \includegraphics[width=1.01\textwidth]{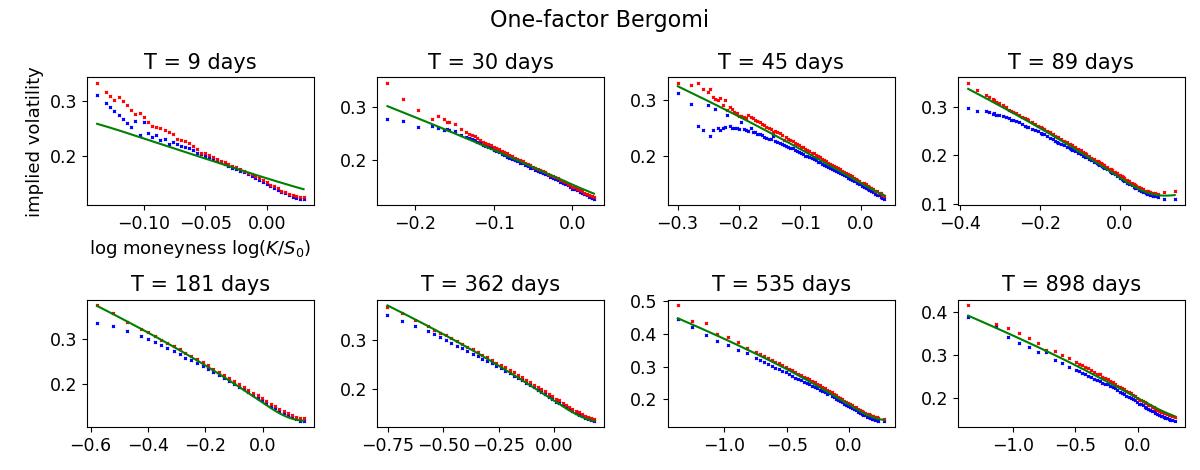}%
    \caption{SPX smiles (bid/ask in blue/red) on 3 July 2013 calibrated by \textbf{one-factor} Bergomi models (green lines).}
  \end{figure}

  \begin{figure}[H]
    \centering    \includegraphics[width=1.01\textwidth]{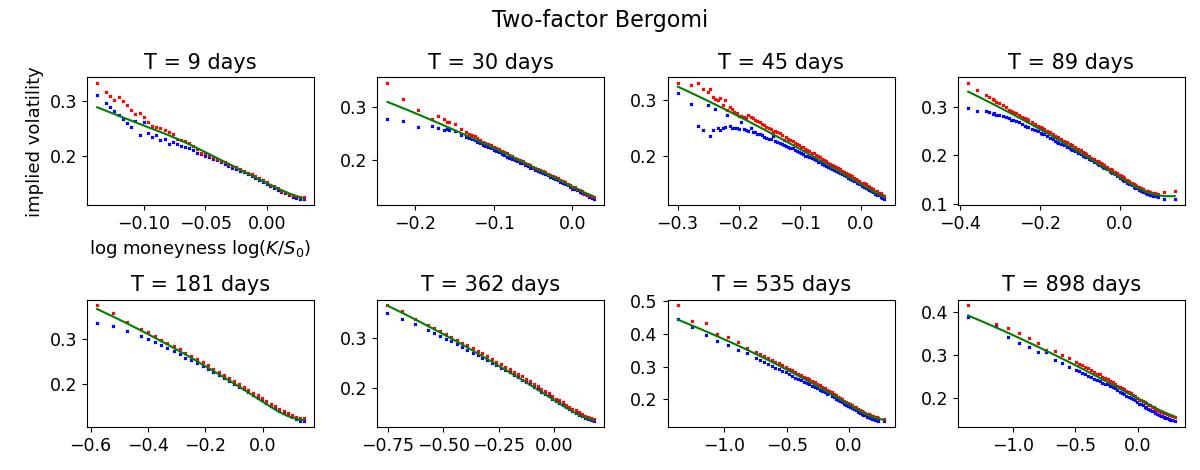}%
    \caption{SPX smiles (bid/ask in blue/red) on 3 July 2013 calibrated by \textbf{two-factor} Bergomi models (green lines).}
  \end{figure}

\underline{October 23, 2017}

  \begin{figure}[H]
    \centering    \includegraphics[width=1.01\textwidth]{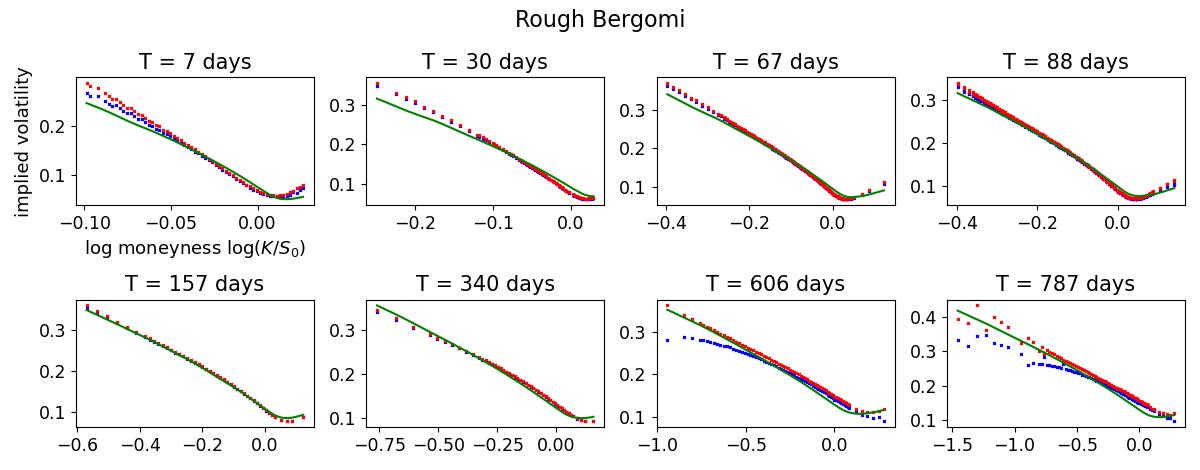}%
    \caption{SPX smiles (bid/ask in blue/red) on 23 October 2017 calibrated by \textbf{rough} Bergomi models (green lines).}
  \end{figure}

  \begin{figure}[H]
    \centering    \includegraphics[width=1.01\textwidth]{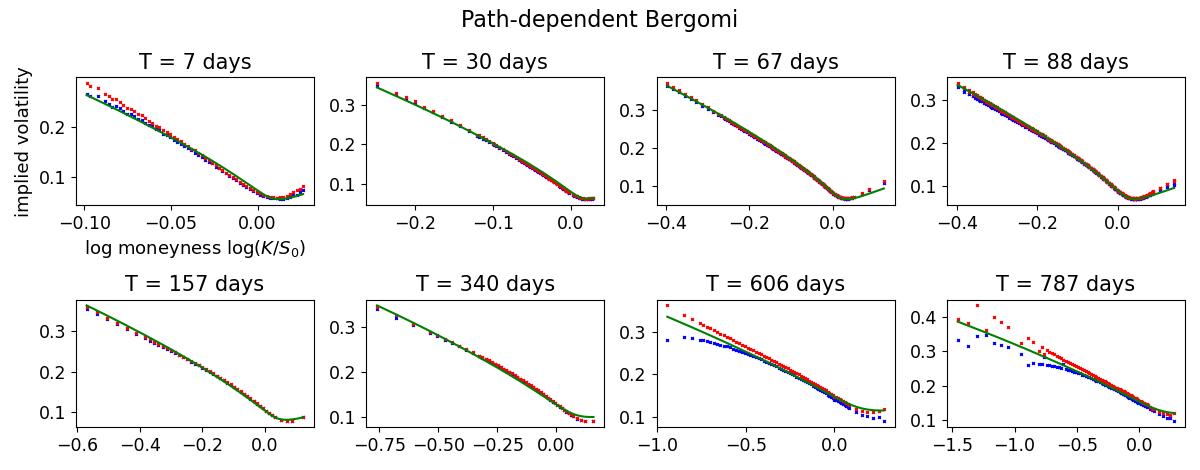}%
    \caption{SPX smiles (bid/ask in blue/red) on 23 October 2017 calibrated by \textbf{path-dependent} Bergomi models (green lines).}
  \end{figure}

  \begin{figure}[H]
    \centering    \includegraphics[width=1.01\textwidth]{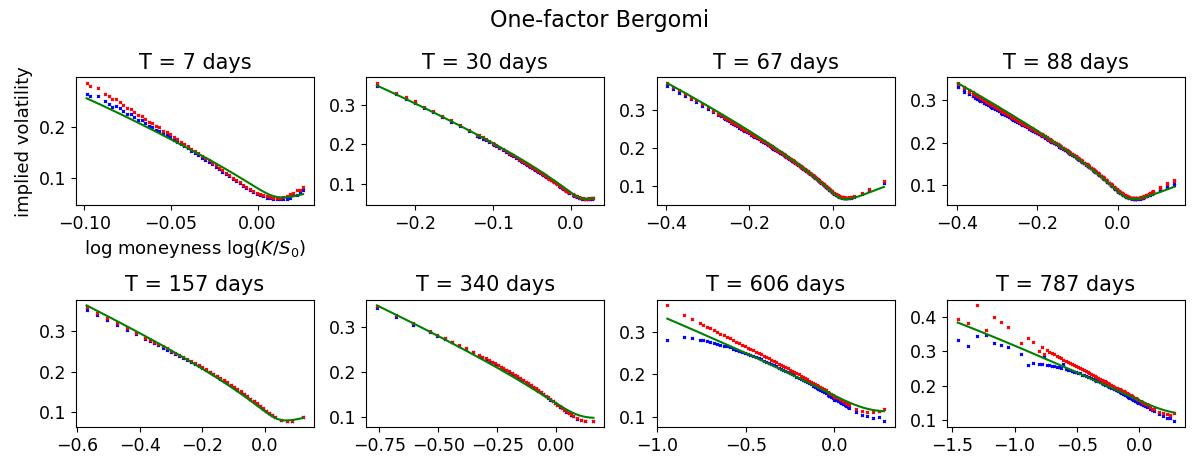}%
    \caption{SPX smiles (bid/ask in blue/red) on 23 October 2017 calibrated by \textbf{one-factor} Bergomi models (green lines).}
  \end{figure}

  \begin{figure}[H]
    \centering    \includegraphics[width=1.01\textwidth]{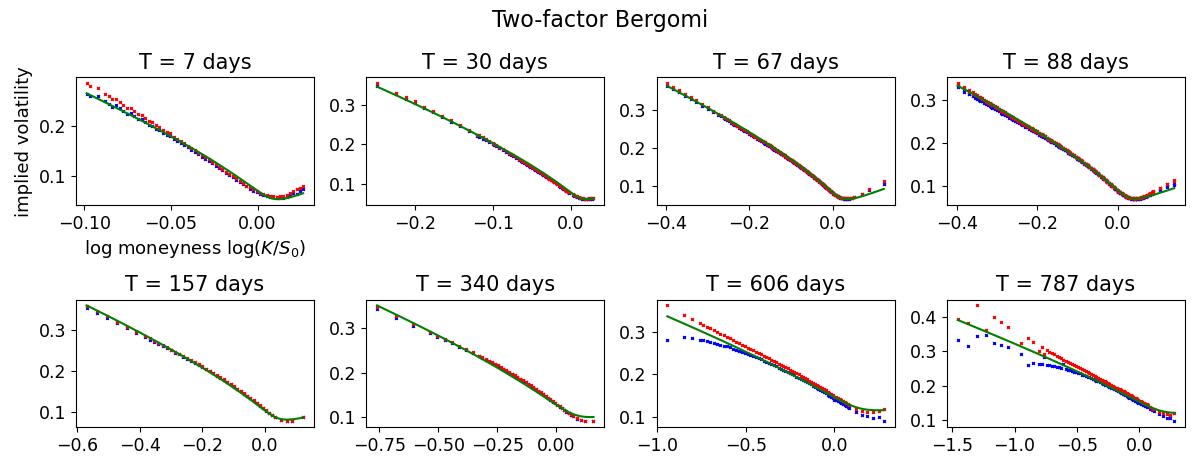}%
    \caption{SPX smiles (bid/ask in blue/red) on 23 October 2017 calibrated by \textbf{two-factor} Bergomi models (green lines).}
  \end{figure}

\section{Fitting the volatility surface: additional graphs}\label{rmse_fit_vs_appendix}

{
\subsection{\textit{Short} maturities}\label{rmse_fit_vs_appendix_3m}

  \begin{figure}[H]
    \centering    \includegraphics[width=0.9\textwidth]{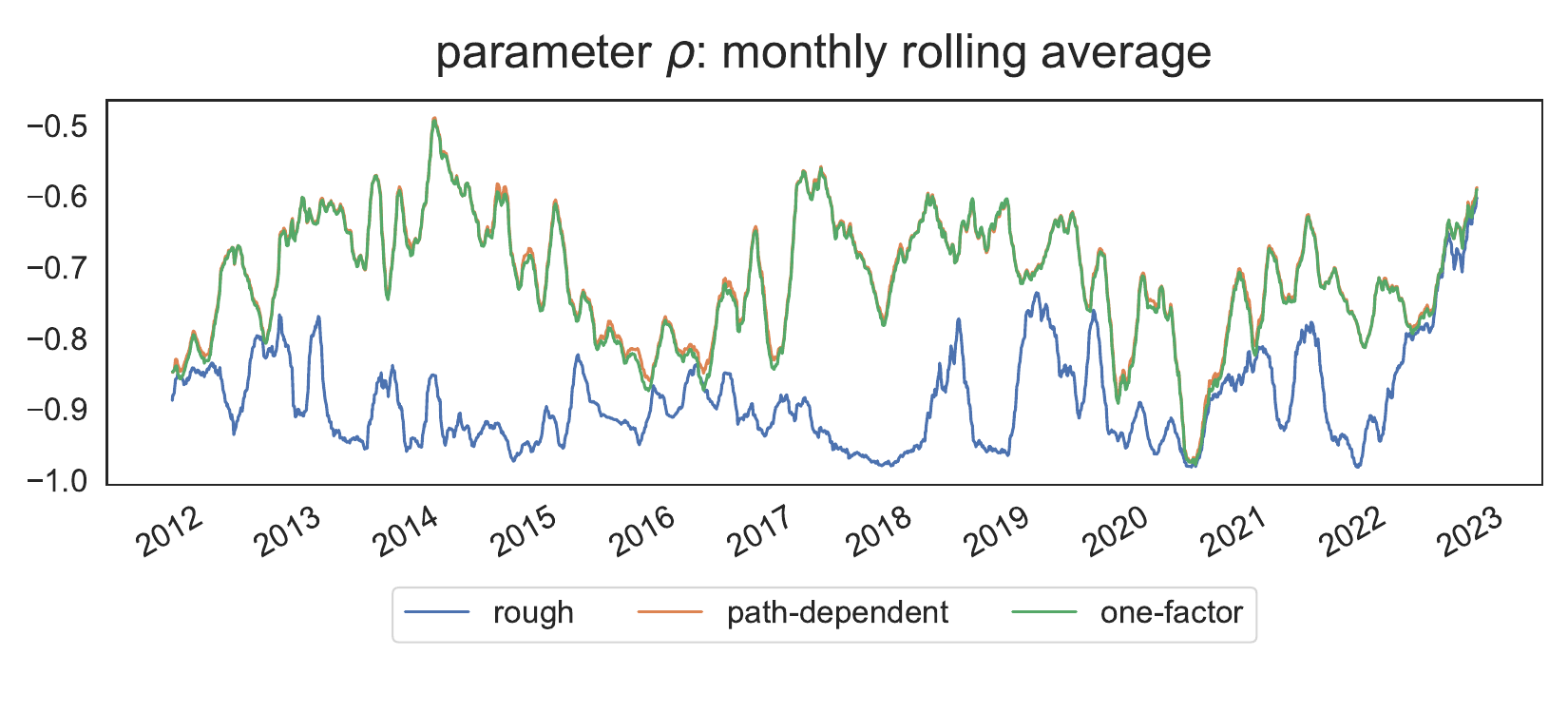}%
    \caption{Time series of monthly average of calibrated parameter $\rho$ for different models calibrated to \textit{short} maturities.}
    \label{fig:calib_rho_3mth}
  \end{figure}

  \begin{figure}[H]
    \centering    \includegraphics[width=0.9\textwidth]{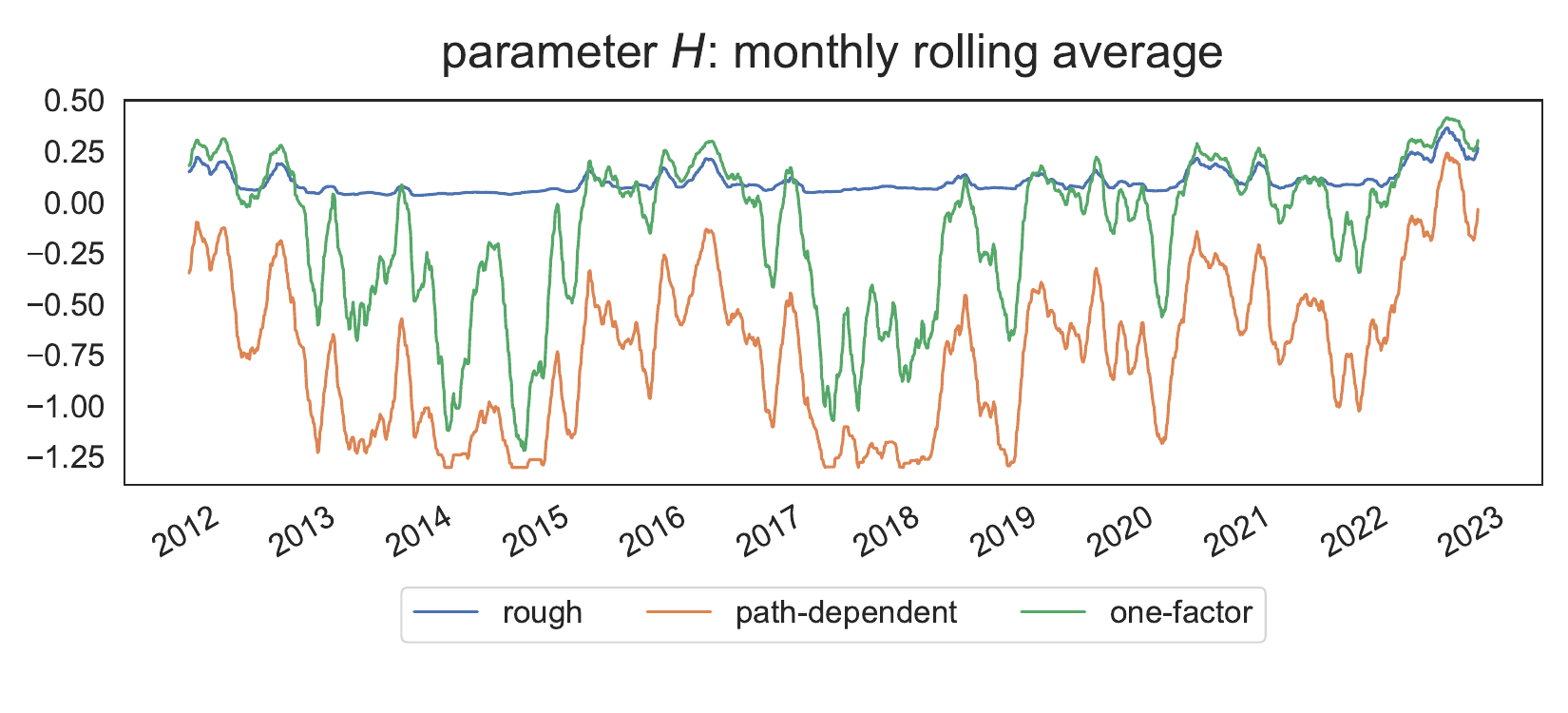}%
    \caption{Time series of monthly average of calibrated parameter $H$ for different models calibrated to \textit{short} maturities.}
    \label{fig:calib_H_3mth}
  \end{figure}
  
  \begin{figure}[H]
    \centering    \includegraphics[width=0.9\textwidth]{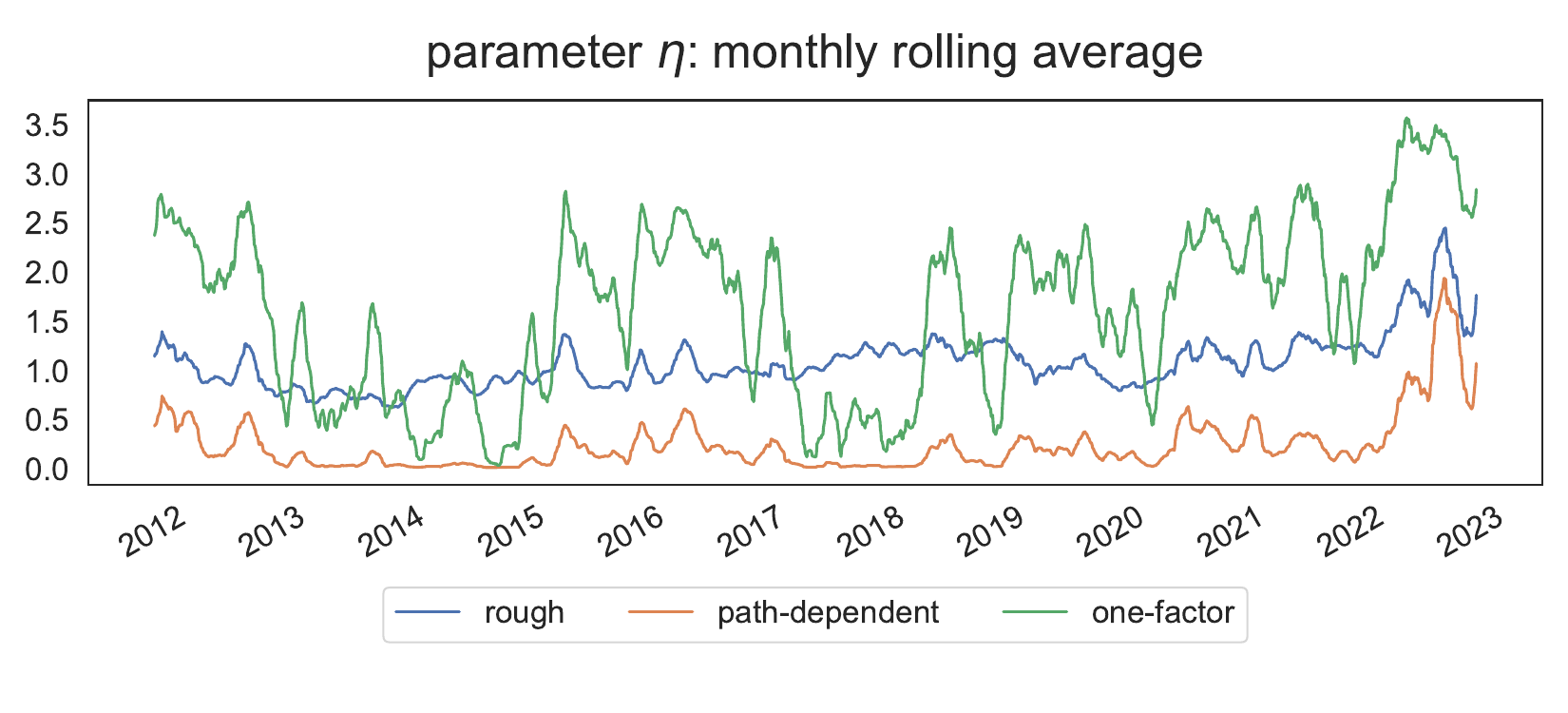}%
    \caption{Time series of monthly average of calibrated parameter $\eta$ for different models calibrated to \textit{short} maturities.}
    \label{fig:calib_eta_3mth}
  \end{figure}

\subsection{\textit{Short and long} maturities}\label{rmse_fit_vs_appendix_3y}

  \begin{figure}[H]
    \centering    \includegraphics[width=0.9\textwidth]{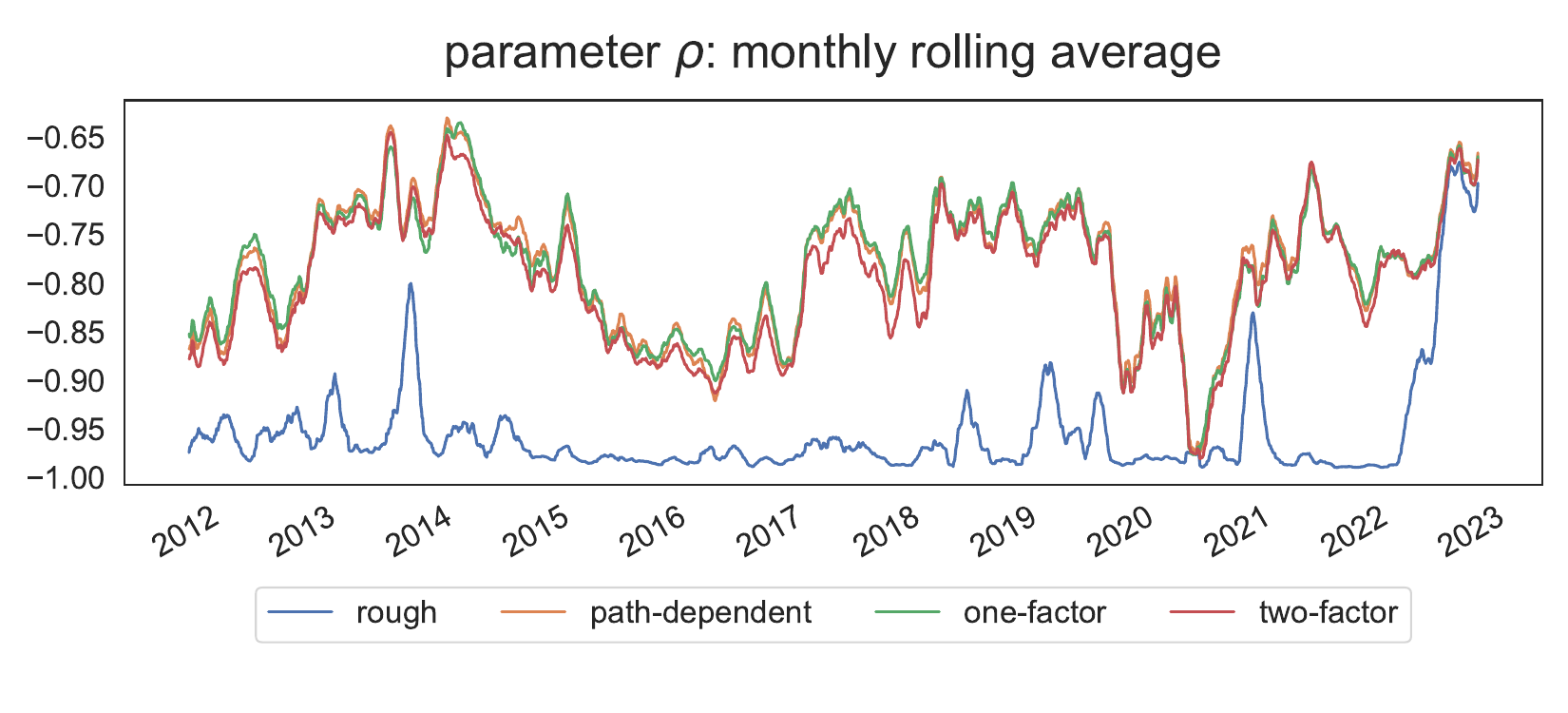}%
    \caption{Time series of monthly average of calibrated parameter $\rho$ of different Bergomi models calibrated to \textit{short and long} maturities.}
    \label{fig:calib_rho}
  \end{figure}

  \begin{figure}[H]
    \centering    \includegraphics[width=0.9\textwidth]{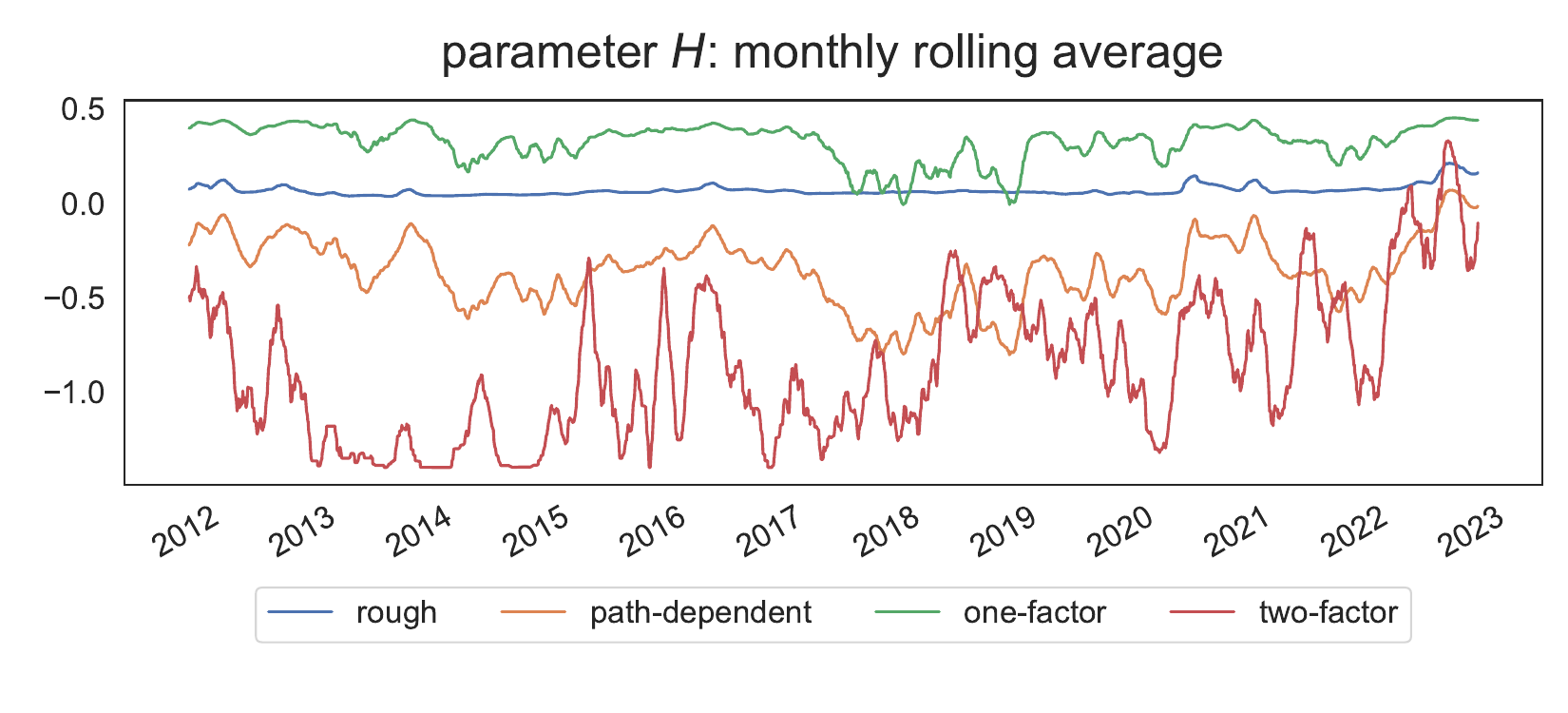}%
    \caption{Time series of monthly average of calibrated parameter $H$ of different Bergomi models calibrated to \textit{short and long} maturities.}
    \label{fig:calib_H}
  \end{figure}

  \begin{figure}[H]
    \centering    \includegraphics[width=0.95\textwidth]{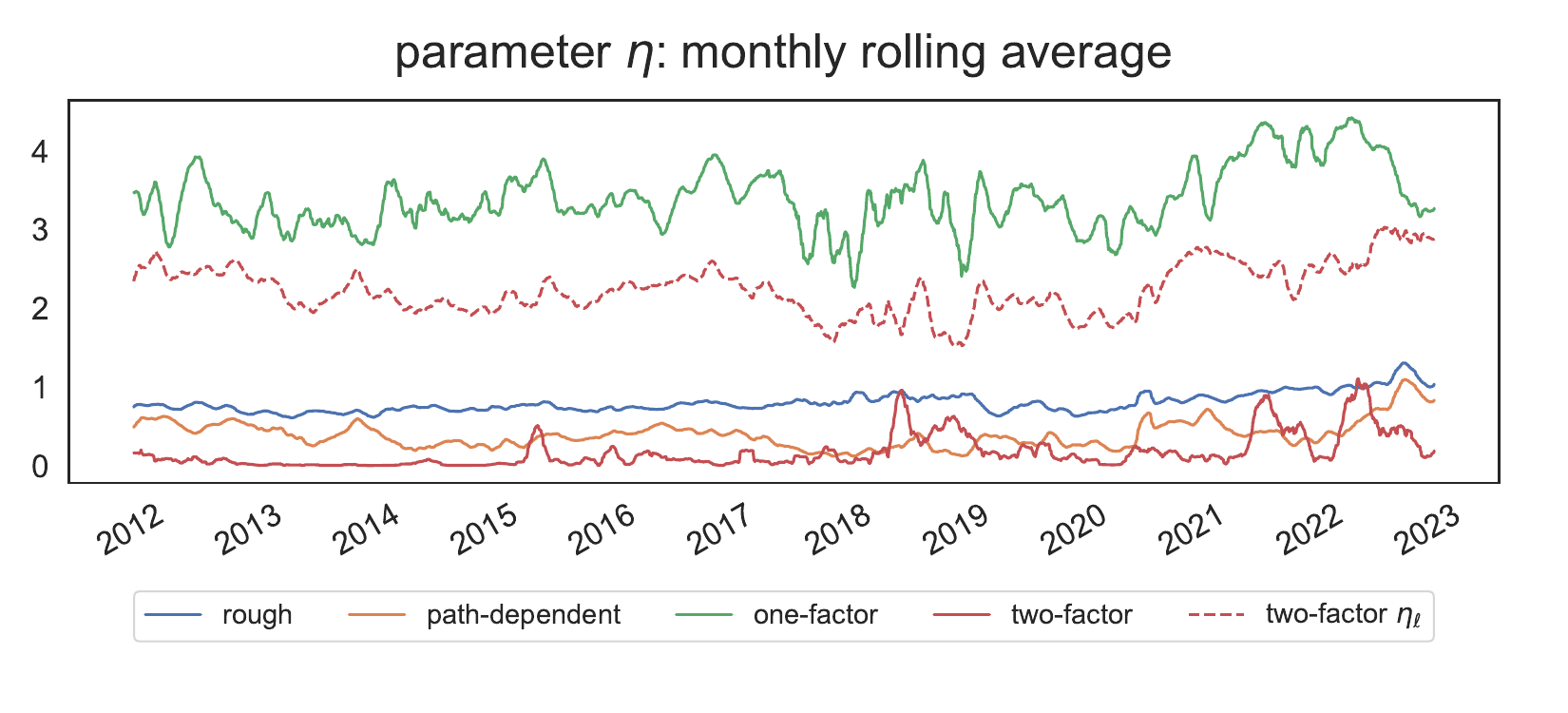}%
    \caption{Time series of monthly average of calibrated parameter $\eta$ of different Bergomi models calibrated to \textit{short and long} maturities.}
    \label{fig:calib_eta_3yr}
  \end{figure}}

\section{Fitting the ATM skew: additional graphs} \label{rmse_fit_skew_appendix}

{

\subsection{\textit{Short} maturities}\label{rmse_fit_skew_appendix_3m}

  \begin{figure}[H]
    \centering    \includegraphics[width=0.8\textwidth]{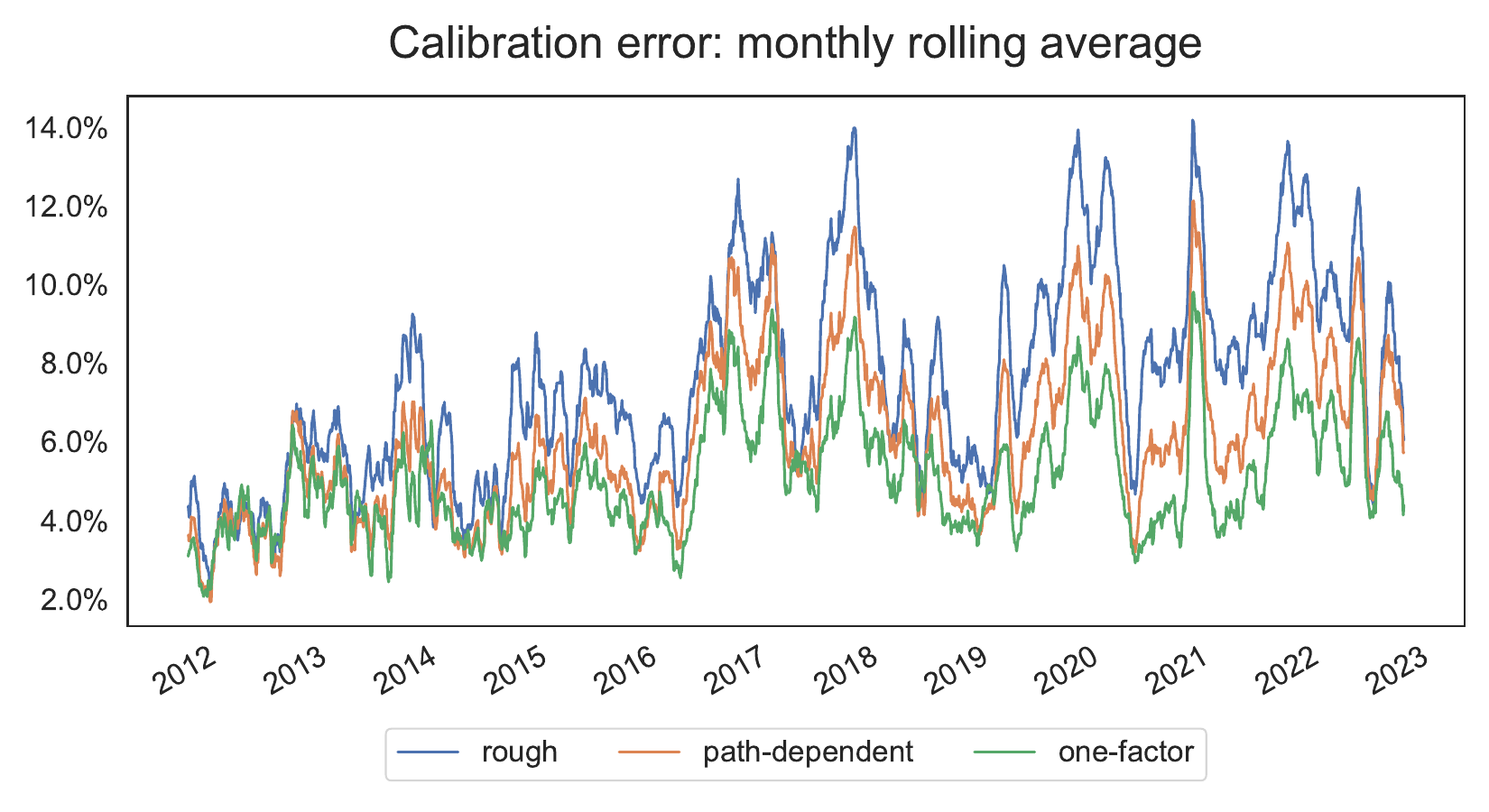}%
    \caption{Time series of monthly rolling average of calibration RMSE of the ATM skew between different Bergomi models for \textit{short} maturities.}
    \label{fig:rmse_3month_skew}
  \end{figure}

{\begin{table}[H]
   \centering  
            \begin{adjustbox}{width=12cm,center}
		\resizebox{\textwidth}{!}{\begin{tabular}{c c c c c c c c} 
	\hline
	\textbf{Model}&\textbf{mean}&\textbf{std}&\textbf{min}&\textbf{5\%}&\textbf{50\%}&\textbf{95\%}&\textbf{max}\\
				\hline 
	\textbf{rough} & 0.0759&	0.0418&	\textbf{0.0005}&	0.0107&	0.0733&	0.1473&	\textbf{0.2912} \\
 	\textbf{path-dependent} & 0.0613&	0.0379&	0.0007&	0.0092&	0.0565&	0.1285&	0.3703\\
 	\textbf{one-factor} & \textbf{0.0508}&	\textbf{0.0330}&	0.0006&	\textbf{0.0058}&	\textbf{0.0466}&	\textbf{0.1093}&	0.3530\\
 \hline 
		\end{tabular}}
  \end{adjustbox}
\vspace{0.02cm}
        \caption{Statistics on the calibration error of the ATM skew for \textit{short} maturities. The lowest error for each statistical measure is in \textbf{bold}.}
        \label{tb:stats_3month_skew} 
\end{table}}

  \begin{figure}[H]
    \centering    \includegraphics[width=0.9\textwidth]{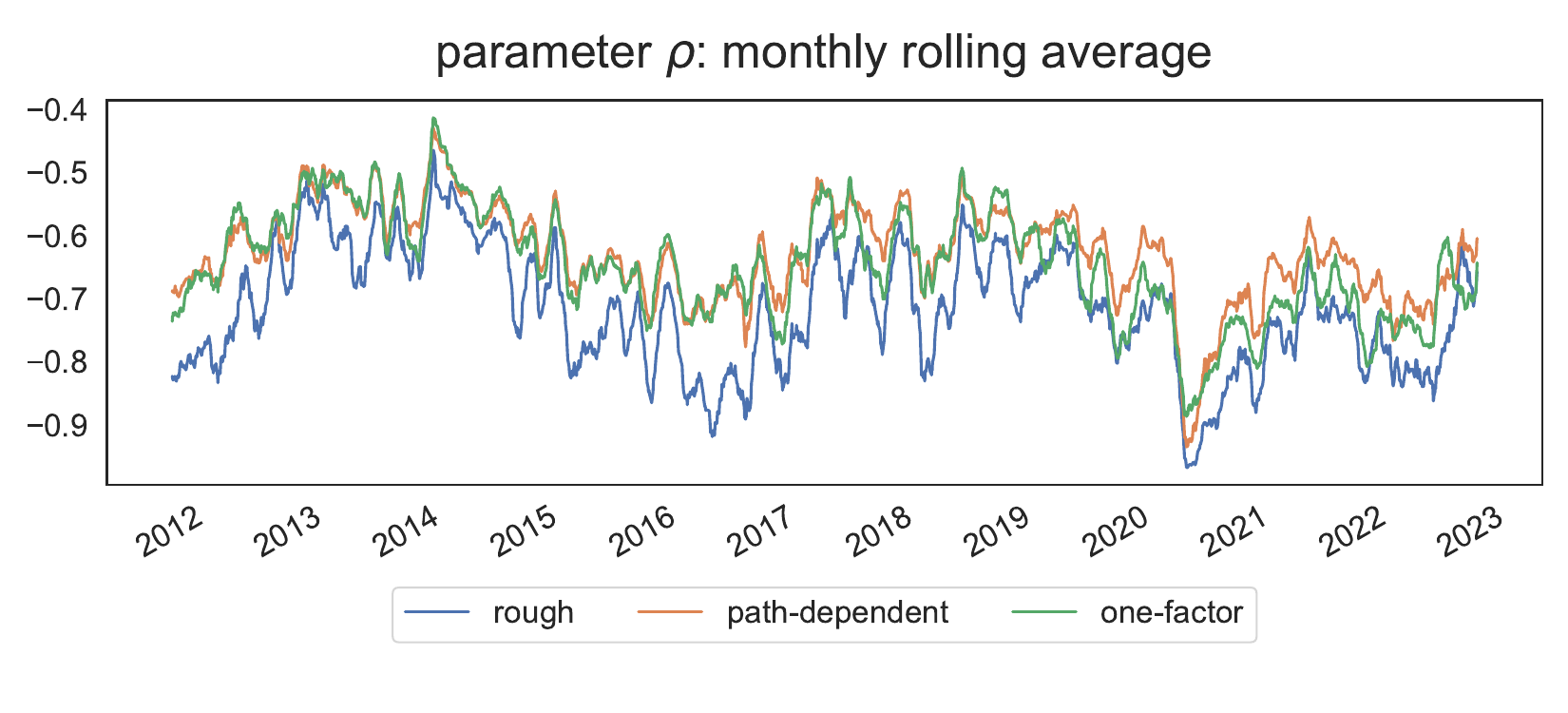}%
    \caption{Time series of monthly average of calibrated parameter $\rho$ for different models calibrated to \textit{short} maturities.}
    \label{fig:calib_rho_3mth_skew}
  \end{figure}

  \begin{figure}[H]
    \centering    \includegraphics[width=0.95\textwidth]{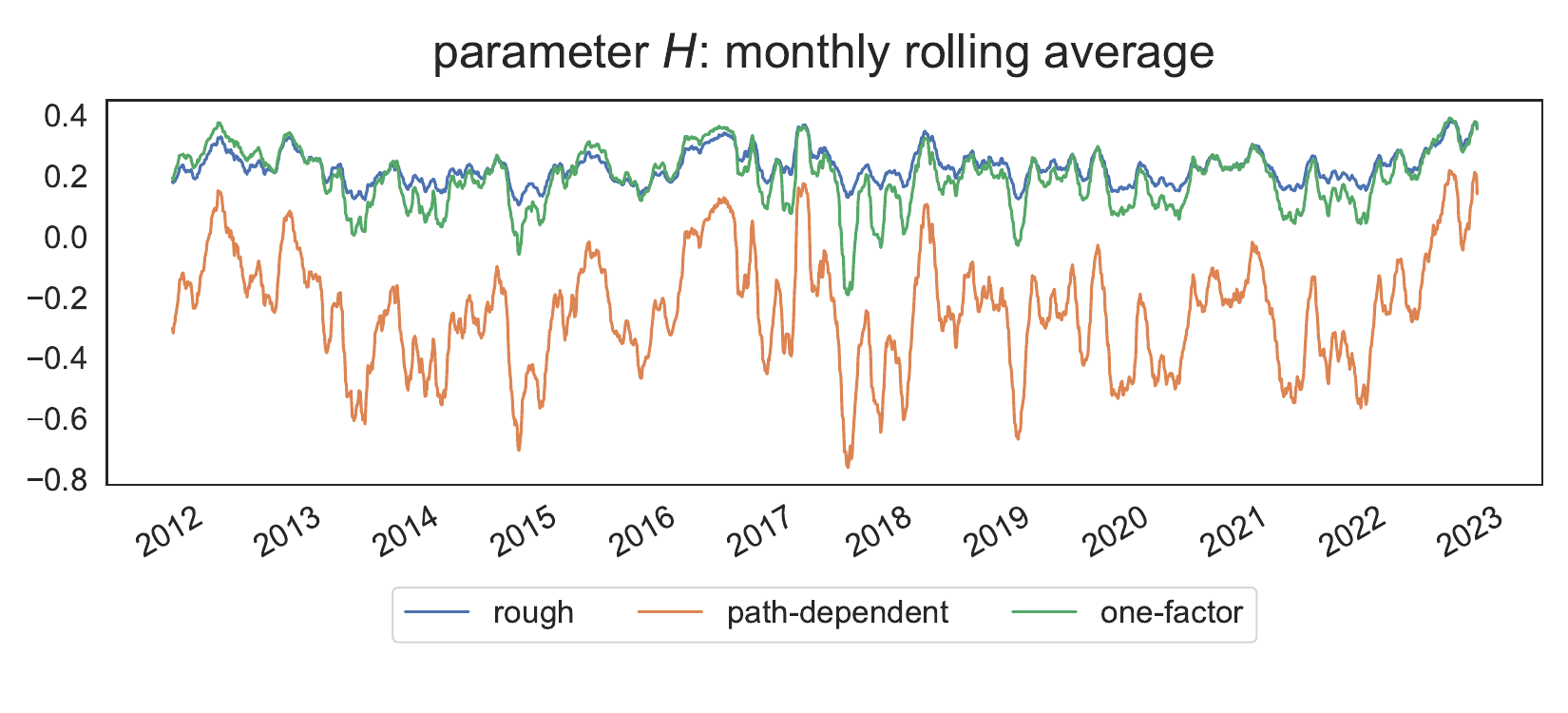}%
    \caption{Time series of monthly average of calibrated parameter $H$ of different Bergomi models calibrated to \textit{short} maturities.}
    \label{fig:calib_H_3ymth_skew}
  \end{figure}

  \begin{figure}[H]
    \centering    \includegraphics[width=0.95\textwidth]{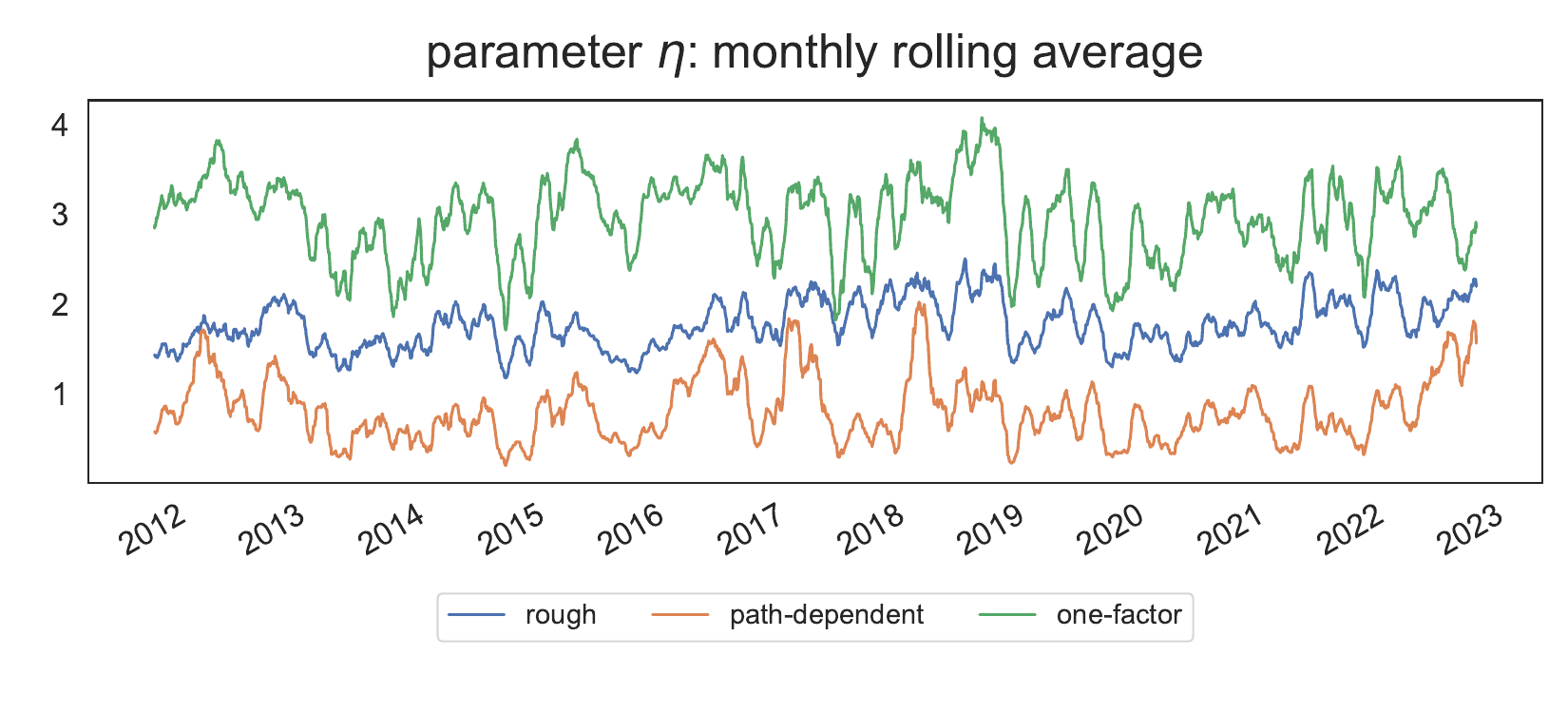}%
    \caption{Time series of monthly average of calibrated parameter $\eta$ of different Bergomi models calibrated to \textit{short} maturities.}
    \label{fig:calib_eta_3mth_skew}
  \end{figure}

\subsection{\textit{Short and long} maturities}\label{rmse_fit_skew_appendix_3y}

  \begin{figure}[H]
    \centering    \includegraphics[width=0.8\textwidth]{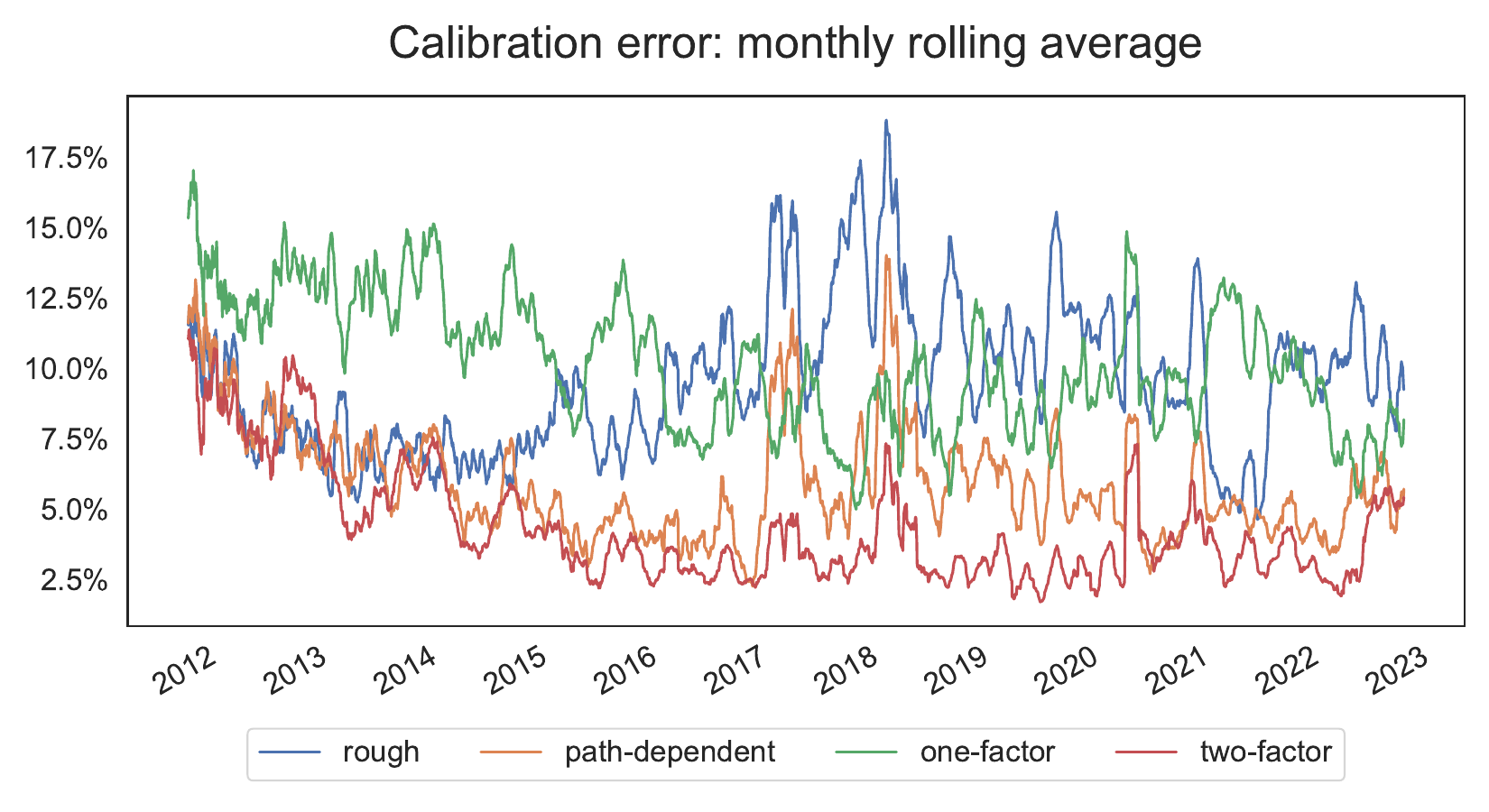}%
    \caption{Time series of monthly rolling average of calibration RMSE of the ATM skew between different Bergomi models for \textit{short and long} maturities.}
    \label{fig:rmse_2year_skew}
  \end{figure}

{\begin{table}[H]
   \centering  
            \begin{adjustbox}{width=12cm,center}
		\resizebox{\textwidth}{!}{\begin{tabular}{c c c c c c c c} 
	\hline
	\textbf{Model}&\textbf{mean}&\textbf{std}&\textbf{min}&\textbf{5\%}&\textbf{50\%}&\textbf{95\%}&\textbf{max}\\
				\hline 
	\textbf{rough} & 0.0955&	0.0473&	0.0022&	0.0379&	0.0877&	0.1749&	0.6947\\
 	\textbf{path-dependent} & 0.0593&	0.0429&	0.0045&	0.0192&	0.0483&	0.1289&	0.6250\\
 	\textbf{one-factor} & 0.1022&	0.0449&	0.0109&	0.0415&	0.0992&	0.1735&	0.6162\\
 	\textbf{two-factor} & \textbf{0.0435}&	\textbf{0.0350}&	\textbf{0.0027}&	\textbf{0.01345}&	\textbf{0.0340}&	\textbf{0.1016}&	\textbf{0.5784}\\
 \hline 
		\end{tabular}}
  \end{adjustbox}
\vspace{0.02cm}
        \caption{Statistics on the calibration error of the ATM skew for the \textit{short and long} maturities. The lowest error for each statistical measure is in \textbf{bold}.}
        \label{tb:stats_3year_skew} 
\end{table}}

  \begin{figure}[H]
    \centering    \includegraphics[width=0.9\textwidth]{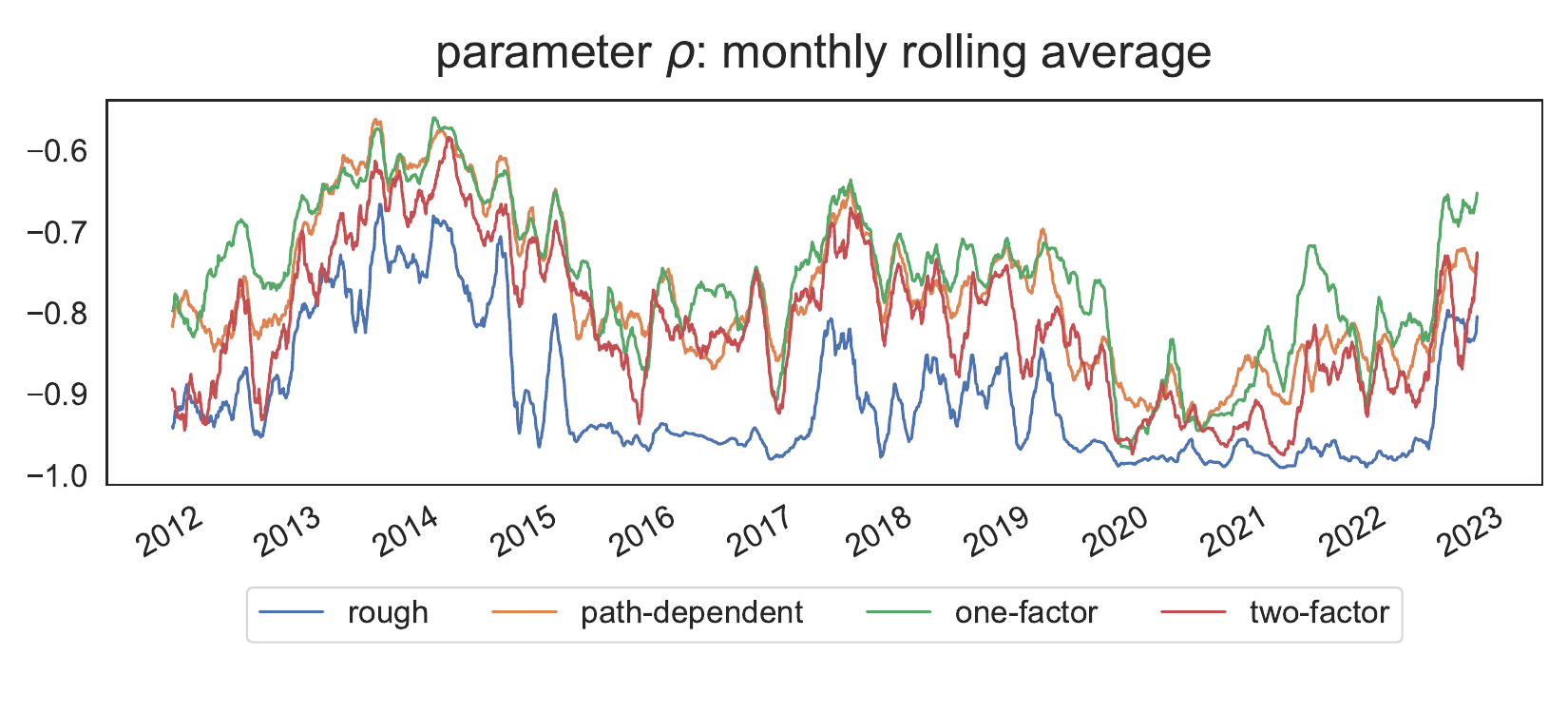}%
    \caption{Time series of monthly average of calibrated parameter $\rho$ for different models calibrated to \textit{short and long} maturities.}
    \label{fig:calib_rho_3yr_skew}
  \end{figure}

  \begin{figure}[H]
    \centering    \includegraphics[width=0.95\textwidth]{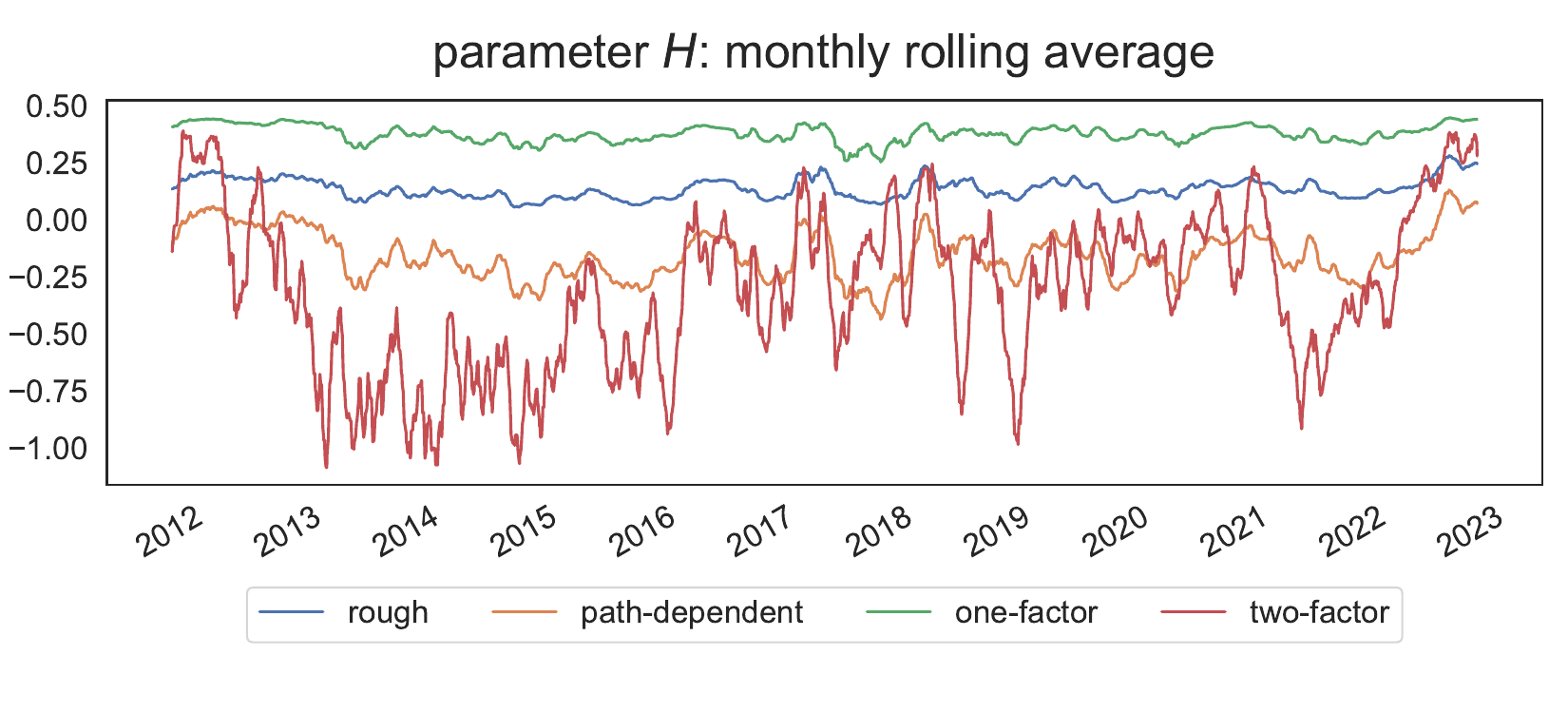}%
    \caption{Time series of monthly average of calibrated parameter $H$ of different Bergomi models calibrated to \textit{short and long} maturities.}
    \label{fig:calib_H_3yr_skew}
  \end{figure}

  \begin{figure}[H]
    \centering    \includegraphics[width=0.95\textwidth]{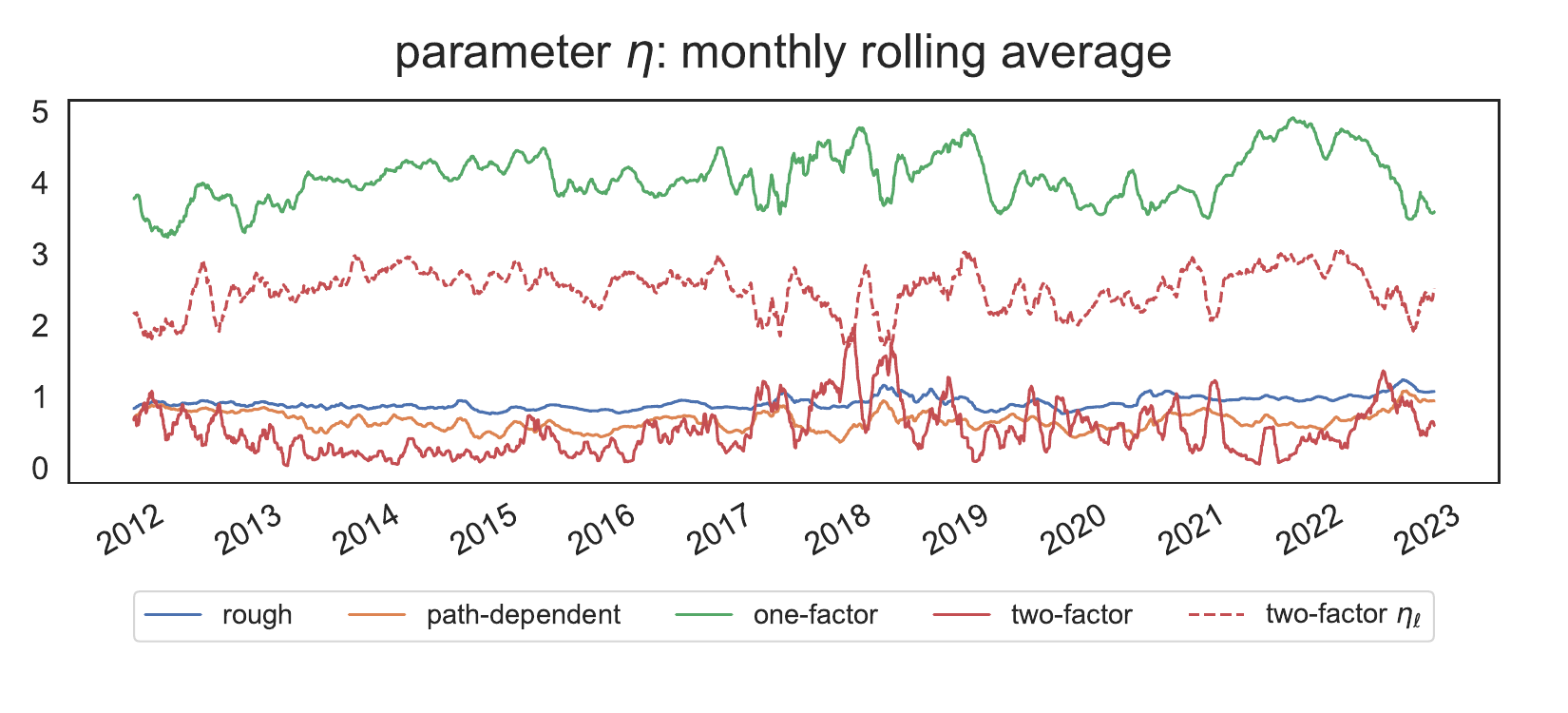}%
    \caption{Time series of monthly average of calibrated parameter $\eta$ of different Bergomi models calibrated to \textit{short and long} maturities.}
    \label{fig:calib_eta_3yr_skew}
  \end{figure}

\section{Sample fits of SPX ATM skew}\label{sample_fit_skew_3years_appendix}
  \begin{figure}[H]
    \centering
    \includegraphics[width=0.5\textwidth]{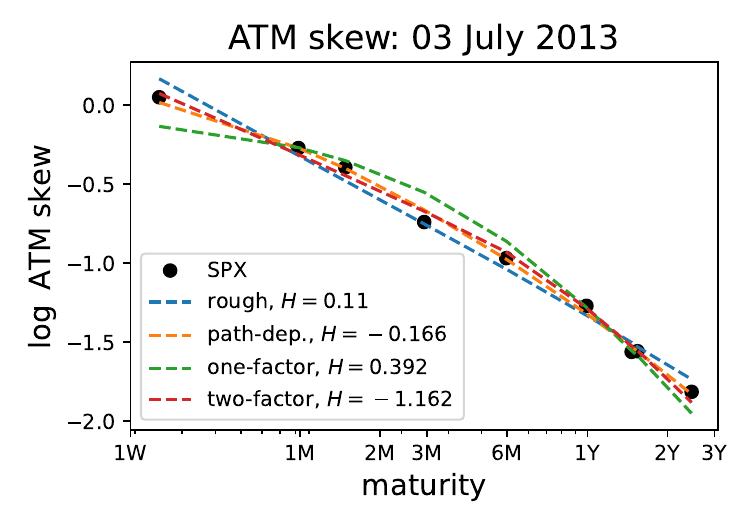}%
    \includegraphics[width=0.5\textwidth]{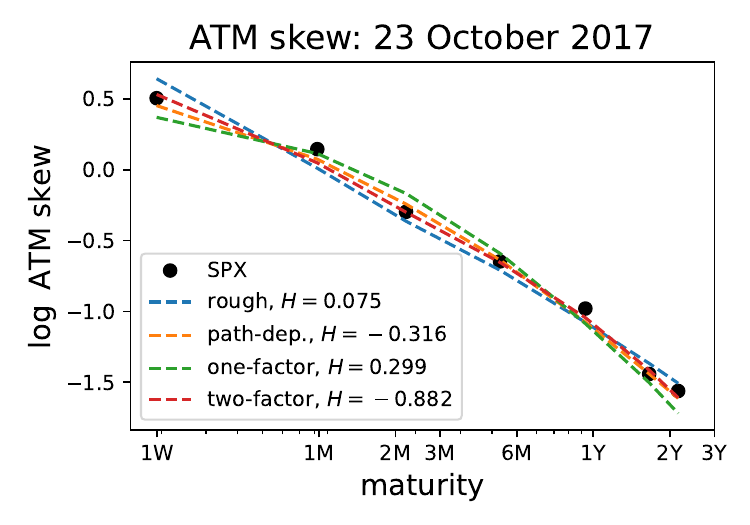}
    \caption{Examples of SPX ATM skew term structure for \textit{short and long} maturities fitted by different Bergomi models in the log-log scale.} 
    \label{fig:fit_skews_3years_examples}
  \end{figure}
}




\bibliographystyle{plainnat}
\bibliography{bibl.bib}

\end{document}